\newcommand{\la}{\langle}
\newcommand{\ra}{\rangle}
\newcommand{\cL}{{\cal L}}
\newcommand{\cO}{{\cal O}}
\newcommand{\nt}{\notag\\}
\newcommand{\q}{\theta}

\newcommand{\cK}{{\cal K}}
\newcommand{\cI}{{\cal I}}
\newcommand{\ep}{\epsilon}
\newcommand{\bq}{\bar\theta}
\newcommand{\da}{{\dot\alpha}}

\renewcommand{\a}{{\alpha}}

\newcommand{\be}{\begin{equation}}
\newcommand{\ee}{\end{equation}}
\newcommand{\bea}{\begin{eqnarray}}
\newcommand{\eaa}{\end{eqnarray}}

\renewcommand{\a}{\alpha}

\newcommand{\cN}{{\cal N}}

\newcommand{\cT}{{\cal T}}

\newcommand{\tr}{{\rm tr}}

\newcommand{\p}[1]{(\ref{#1})}
\newcommand{\bt}[1]{{\bar t}}

\newcommand \vev [1] {\langle{#1}\rangle}

\newcommand\lr[1]{{\left({#1}\right)}}

\documentclass[12pt,a4paper]{article}
\input epsf
\usepackage{amsmath,amsfonts,color,latexsym,epsfig}
\usepackage{amssymb}
\usepackage[english, USenglish]{babel}
\usepackage{psfrag}
\usepackage{a4wide}
\usepackage{rotating}
\usepackage{cite}
\usepackage{youngtab}
\usepackage{array}
\usepackage{hyperref}

\newcommand{\captionfonts}{\small}

\makeatletter  
\long\def\@makecaption#1#2{%
  \vskip\abovecaptionskip
  \sbox\@tempboxa{{\captionfonts #1: #2}}%
  \ifdim \wd\@tempboxa >\hsize
    {\captionfonts #1: #2\par}
  \else
    \hbox to\hsize{\hfil\box\@tempboxa\hfil}%
  \fi
  \vskip\belowcaptionskip}
\makeatother   

\csname @addtoreset\endcsname{equation}{section}
\usepackage{amsmath,a4wide}
\usepackage{graphics,graphicx}

\begin{document}
  
\thispagestyle{empty}


\null\vskip-43pt \hfill
\begin{minipage}[t]{50mm}
CERN-PH-TH/2011-208 \\
DCPT-11/33 \\
IPhT--T11/176 \\
LAPTH--030/11 
\end{minipage}

\vskip1.5truecm
\begin{center}
\vskip 0.2truecm

 {\Large\bf
Hidden symmetry of four-point correlation functions  \\[2mm] and amplitudes in $\cN=4$ SYM }
\vskip 1truecm

\vskip 1truecm
{\bf    Burkhard Eden$^{a}$, Paul Heslop$^{a}$, Gregory P. Korchemsky$^{b}$, Emery Sokatchev$^{c,d,e}$ \\
}

\vskip 0.4truecm
$^{a}$ {\it  Mathematics Department, Durham University, 
Science Laboratories,
 \\
South Rd, Durham DH1 3LE,
United Kingdom \\
 \vskip .2truecm
$^{b}$ Institut de Physique Th\'eorique\,\footnote{Unit\'e de Recherche Associ\'ee au CNRS URA 2306},
CEA Saclay, \\
91191 Gif-sur-Yvette Cedex, France\\
\vskip .2truecm $^{c}$ Physics Department, Theory Unit, CERN,\\ CH -1211, Geneva 23, Switzerland \\
\vskip .2truecm $^{d}$ Institut Universitaire de France, \\103, bd Saint-Michel
F-75005 Paris, France \\
\vskip .2truecm $^{e}$ LAPTH\,\footnote[2]{Laboratoire d'Annecy-le-Vieux de Physique Th\'{e}orique, UMR 5108},   Universit\'{e} de Savoie, CNRS, \\
B.P. 110,  F-74941 Annecy-le-Vieux, France
                       } \\
\end{center}

\vskip -.2truecm 

\centerline{\bf Abstract}
\medskip
\noindent 
{
We study the four-point correlation function of stress-tensor
supermultiplets in $\cN=4$ SYM using the method of
Lagrangian insertions. We  argue that, as a corollary of $\cN=4$ superconformal 
symmetry,  the resulting all-loop
integrand possesses an unexpected complete symmetry under the exchange of the four external and all the internal (integration)
points. This alone allows us to predict the integrand of the three-loop correlation function up to four undetermined constants. Further, exploiting the 
conjectured amplitude/correlation function duality, we are able
to fully determine the three-loop integrand in the planar limit.  
We perform an independent check of this
result by verifying that it is consistent with
the operator product expansion, in particular that it correctly reproduces the
three-loop anomalous dimension of the Konishi operator.  As a
byproduct of our study, we also obtain the three-point function of two half-BPS operators  and one Konishi operator at three-loop level. 
We use the same technique to work out a compact form for the four-loop four-point integrand
and discuss the generalisation  to higher loops.

}

\newpage

\setcounter{page}{1}\setcounter{footnote}{0}

\section{Introduction}

In the first few years after the discovery of the AdS/CFT
correspondence~\cite{123} considerable effort was put into
studying gauge-invariant operators and their  correlation functions
in perturbative $\cN=4$ SYM, see for
example~\cite{oneTwo,ESS,Rome,partialNonRen,EHW,Heslop:2001dr,Heslop:2003xu,Heslop:2002hp,Dolan:2004mu,Dolan:2001tt}.

In parallel a great deal of progress has been made in
exploring seemingly
completely different objects in  $\cN=4$ SYM, namely scattering
amplitudes. One of the important results in this arena has
been the discovery that {planar} MHV amplitudes can be described by null
polygonal Wilson loops~\cite{aldMald,dual11,dual12}. This has led to
much progress in understanding the
amplitudes themselves (since so far the Wilson loop integrals have
been easier to compute than the 
corresponding amplitude
integrals~\cite{dual21,dual22,Anastasiou:2009kna,DelDuca:2010zg,DelDuca:2010zp,spradGonch,Heslop:2010kq}),
culminating in the discovery of a new symmetry, dual conformal
symmetry~\cite{dual11} and
dual superconformal symmetry~\cite{annecySuperspace,BM,Brandhuber:2008pf}. This in turn has been instrumental in recent
progress in studying scattering amplitudes at the level of the integrand 
beyond the MHV
case~\cite{nima1,nima2} and in the construction of a supersymmetric  Wilson
loop~\cite{marksAndSparks,simon}, which should reproduce the {super}-amplitude {in planar $\cN=4$ SYM}
provided the correct regularisation is understood~\cite{belitsky}.

{Excitingly, a year ago it was discovered that correlation functions
of gauge-invariant
operators, defined as superconformal primaries of $\cN=4$ supermultiplets,  describe both Wilson loops~\cite{AEMKS} and
MHV amplitudes~\cite{EKS2}. These new dualities require taking the light-like (or ``on-shell") limit in which the operators are located at 
the vertices of a null polygon. The  correlation function/Wilson loop
duality can be  viewed as a relation between the (properly
regularised) integrals in terms of which both objects are expressed.
However,  the correlation function/amplitude duality is {most easily} understood at the level of the {\it integrand}~\cite{EKS3}. The natural definition of
the integrand for loop-level correlation functions via the Lagrangian insertion procedure
remarkably reproduces the amplitude integrand derived in momentum
twistor space~\cite{nima2}. Both the MHV amplitude and the correlation function
of superconformal primaries have natural superspace generalisations, and
indeed (if we switch off the anti-chiral half of the superspace coordinates), the full
supercorrelation functions of the supermultiplets continue to reproduce all non-MHV
superamplitude integrands~\cite{paper1,paper2,mscorrelationfunctions}.\footnote{
We should note that it is expected that this duality holds for correlation functions of a large
class of operators, 
but so far has only been checked for the stress-tensor
multiplet~\cite{paper1,paper2} and 
the Konishi multiplet~\cite{AEMKS,mscorrelationfunctions}.
An
explanation for this duality was proposed  in momentum twistor
space~\cite{mscorrelationfunctions}. However this explanation is at  a rather 
formal level due to the regularisation issue alluded to earlier, so 
at the moment we view the duality as conjectural.}}

{We expect that this duality will lead to new insights both for
correlation functions and amplitudes, and it is the purpose of this
paper to give an example.  Namely, we shall exploit
the duality, combined with a new symmetry to be described below, to 
derive the correlation function of four  stress-tensor supermultiplet  operators  at
three and four loops and to set up a procedure for going to higher loops. }

The stress-tensor multiplet operator $\cT$ is the simplest example of the so-called half-BPS multiplets. Its lowest-weight state (superconformal primary)  is a scalar operator $\cO_{\mathbf{20'}}$ of scaling dimension two in the $\mathbf{20'}$ of $SU(4)$.   In the context of the AdS/CFT correspondence the half-BPS operators are regarded as the duals of the AdS$_5\times S^5$ supergravity states. They have the important property that their two- and three-point functions are protected from quantum corrections. The first non-trivial example of a correlation function of such multiplets is provided by the four-point case. In the past the four-point functions of the operators $\cO_{\bf 20'}$ have been extensively studied, both at strong coupling (in AdS supergravity) \cite{ADS} and at weak coupling \cite{oneTwo,ESS,Rome,partialNonRen,EHW,Heslop:2001dr,Heslop:2003xu,Heslop:2002hp,Dolan:2004mu,Dolan:2001tt}. The perturbative calculations used the standard Feynman diagram technique adapted to $\cN=1$ superspace \cite{Rome} or to $\cN=2$ harmonic superspace \cite{ESS}. Beyond two loops this technique becomes exceedingly difficult, so no explicit three-loop results for four-point functions are available. Our aim in this paper is not to carry out a Feynman diagram calculation, but to {\it predict} the form of the correlation function in terms of a set of scalar Feynman integrals.

There are two main ingredients which allow us to do this. Firstly, we discover a hidden permutation symmetry of the integrand of the four-point correlation function, which exchanges the integration points  and the external points. This symmetry can be seen in the framework of the Lagrangian insertion procedure. The $\ell-$loop correction to the four-point correlation function of the scalar operators $\cO_{\mathbf{20'}}$ is given by their $(4+\ell)-$point Born-level correlation function with $\ell$ Lagrangian insertions. The $\cN=4$ SYM Lagrangian is a member of the same stress-tensor supermultiplet $\cT$ as the operator  $\cO_{\mathbf{20'}}$ itself.  Consequently, what determines the integrand of the four-point $\ell-$loop
correlation function is the Born-level correlation function of $(4+\ell)$ operators $\cT$.  It can be described by a nilpotent
superconformally covariant $(4+\ell)-$point polynomial in $\cN=4$ analytic superspace \cite{paulN4,Heslop:2003xu}. Our new observation is that this nilpotent object has full permutation $S_{4+\ell}$ symmetry. This, together with the crossing symmetry of the correlation function of the super-operators $\cT$, lead to the aforementioned new symmetry of the four-point integrand. 

This hidden symmetry, together
with reasonable assumptions about the pole  structure of the
integrand following from the OPE, already yield a very constrained form for the four-point correlation function
integrand at any loop level. At three loops, it determines the result, for an arbitrary gauge group (e.g.,  $SU(N_c)$ for any value of $N_c$),  up to four unknown
coefficients.

The second main ingredient in our construction is the correlation function/amplitude duality. According to it, the integrand of the {\it planar} four-point correlation function of protected operators $\cO_{\mathbf{20'}}$, in the light-like limit where the four points become sequentially light-like separated,   is equal to the square of the integrand of the {\it planar} four-gluon scattering amplitude. We use this duality to compare the most general form of the four-point correlation function described above with the known expression for the three-loop four-gluon 
amplitude~\cite{bern42}. We find
that, for a certain choice of the four coefficients, we get precise
agreement, thus at the same time determining the unique form of the
planar correlation function at three loops and providing further non-trivial
evidence for the duality. It is interesting to notice that, besides the known three-loop integrals of ``ladder" and ``tennis court" type, we find only two new three-loop planar integral topologies. 

Our procedure is not limited to three loops. It is rather straightforward to use the results on the four-gluon amplitude at four \cite{4loopMHV} and even five \cite{5loops} loops and predict the form of the integrand of the correlation function. In this paper we treat the four-loop case and obtain a compact
expression for the correlation function integrand there.  

{ 
Although in this paper our focus is on the correlation functions, we
wish to emphasise that the
techniques have much to say about the amplitude integrands
as well. In particular the new permutation symmetry, which relates external
points and integration points, is also present for the amplitudes (albeit in
a broken form). This together with the fact that the amplitude/correlation
function duality relates products of lower-loop amplitudes to
the correlation function means that a large portion of the amplitude
can be found from lower-loop amplitudes. For example, the entire three-loop
amplitude can be determined from the one- and two-loop amplitudes and a large
portion of the four-loop amplitude as well. We comment further on this point
in the conclusions.
}

The paper is organised as follows. {In Section \ref{sec:gen} we 
define the correlation functions of the stress-tensor multiplet and give a short
review of the Lagrangian insertion technique.}
 In Section~\ref{sec:three-loop-4} we derive the three-loop correlation function and
check the agreement with the {four-gluon scattering amplitude in the light-like limit}. In Section~\ref{sec:ope-test} we
test the correlation function by an OPE analysis  and we show that it correctly reproduces the known value for the three-loop Konishi
anomalous dimension \cite{klovUs}.   This section also contains our
prediction for the three-point correlation function of two protected
scalar operators and the Konishi operator at three loops. 
In Section~\ref{sec:four-loops} we derive a
similar expression at four loops albeit without the benefit of the
analogous OPE check.  
{In the conclusions we comment on the implications of these techniques
for amplitudes and also discuss the conditions for planarity.}   
In Appendix~\ref{sec:an-s_n-invariant} we
prove the hidden permutation symmetry which is the key to the simple
form we find for the three-loop correlation function. It is due to the
existence of an $S_n-$invariant, 
conformally covariant, nilpotent polynomial which we give a 
simple expression for here. Appendix~\ref{sec:three-loop-correlation function} contains some technical details on the numerical evaluation of the integrals needed for the OPE test.

\section{Correlation functions of
the stress-tensor multiplet in $\cN=4$ SYM}
\label{sec:gen}

In this section, we define the correlation functions of
the stress-tensor multiplet in the $\cN=4$ SYM theory and discuss 
their general form. We begin by presenting the notation 
(which is inherited from~\cite{paper1,paper2}) and summarise some
basic results concerning correlation functions in $\cN=4$ SYM, following from $\cN=4$ superconformal symmetry.  

\subsection{The $\cN=4$ stress-tensor multiplet in analytic superspace}

The field content of the $\cN=4$ super-Yang-Mills theory comprises six
real scalars $\Phi^I$ (with $I=1,\ldots,6$ being an $SO(6)$ index), four Weyl fermions (and their complex conjugates) 
and the gauge field, all in the adjoint representation of, e.g., the $SU(N_c)$ gauge group. 
In this paper, we study the correlation functions of the special class 
of the so-called half-BPS gauge-invariant local operators. One of the important properties 
of such operators is that their scaling dimension is protected from perturbative corrections. 

The simplest example of a half-BPS operator is 
the bilinear gauge-invariant operator made of the six scalars,
\begin{align}\label{O-IJ}
  \mathcal{O}_{\mathbf{20'}}^{IJ} &= \tr\left(\Phi^I \Phi^J\right) - \frac16 \delta^{IJ} \tr\left(\Phi^K \Phi^K\right)\,.
\end{align}
The operator $\mathcal{O}^{IJ}$ belongs to the representation 
$\mathbf{20'}$ of the $R$ symmetry group $SO(6)\sim SU(4)$. It is the lowest-weight state 
of the so-called $\cN=4$ stress-tensor supermultiplet   containing, among others, the stress-tensor (hence the name) and the Lagrangian of the theory. The latter appear as the coefficients in the expansion of
half-BPS superfield operator $\cT(x,\theta^A,\bar \theta_A)$ in powers of
the odd variables $\theta^A$ and $\bar\theta_A$ (with $A=1,\ldots,4$ being an $SU(4)$ index). 

The half-BPS superfield $\cT(x,\theta^A,\bar \theta_A)$ satisfies constraints implying that it depends on half of the odd variables, both chiral and anti-chiral. \footnote{In this half-BPS superfields radically differ from chiral superfields $F(x,\q^A)$ which depend only on the chiral half of the odd variables.} The appropriate formalism which makes the $\cN =4$ half-BPS property manifest is  that of 
$\cN =4$ analytic superspace \cite{paulN4} or, equivalently, $\cN =4$ harmonic superspace \cite{Andrianopoli:1999vr}.  On the other hand, the lack of an off-shell formulation of $\cN = 
4$ SYM  makes $\cN =4$ analytic superspace inadequate for 
performing Feynman graph calculations of the correlation functions. These are most conveniently done in $\cN =2$ harmonic superspace \cite{GIKOS}, after which the results can be lifted to $\cN =4$.

Harmonic/analytic superspace makes use of auxiliary bosonic (``harmonic") $y-$variables 
in order to covariantly break the  R symmetry group $SU(4)$ down to $SU
(2)\times SU(2)' \times U(1)$. Specifically, the chiral Grassmann coordinate $\q^A_\a$   can be decomposed into its halves by splitting the $SU(4)$ index $A=(a,a')$
(with $a,a'=1,2$) and substituting
\begin{align}\label{anth}
\q^A_\a \ \rightarrow \ 
(\rho^a_\a \,,   \q^{\,a'}_\a)\,,\quad \text{with} \ \ \rho^a_\a= \q^a_\a + \q^{a'}_\a y^a_{a'}\,,
\end{align}
and similarly for the anti-chiral $\bq^\da_A \ \rightarrow \ ( \bar\rho^\da_{a'}\,,  \bq^\da_a)$. 
Here the harmonic variables $y^a_{a'}$ carry $SU(2)\times SU(2)'$  indices and  parametrise the four-dimensional  coset  $SU(4)/(SU(2)\times SU(2)' \times U(1))$  (or its complexification in 
the approach of \cite{paulN4}).  The projected odd variables carry also a $U(1) \subset SU(4)$ weight $(+1)$ for $\rho$ and $(-1)$ for $\bar\rho$. Then, the stress-tensor supermultiplet is defined as a function on {\it analytic superspace}    
\begin{align}\label{TT}
\cT=\cT( {x^{\dot \alpha\a}}, \, {\rho^a_\a },
\, \bar \rho_{\, a'}^{\dot \alpha}, \, y_{a'}^a)\,.
\end{align}
The half-BPS nature of $\cT$ manifests itself in the independence of this function
on $\q^{\,a'}_\a$ and $\bq^\da_a$. This is the meaning of the so-called ``BPS shortening"  of the stress-tensor supermultiplet. 

The supermultiplet  (\ref{TT}) carries $U(1)$ weight $(+4)$ and its scaling dimension is protected from quantum corrections. Its lowest-weight component (superconformal primary) is given by the bilinear scalar operator (\ref{O-IJ}) 
\begin{align}\label{defO}
\cO(x,y) =Y_I \, Y_J\, \mathcal{O}_{\mathbf{20'}}^{IJ}(x) =Y_I \, Y_J\,  \tr\left(\Phi^I \Phi^J\right)  \,, 
\end{align}
where 
the $SO(6)$ indices have been projected with the $SO(6)$ harmonic variable $Y_I=Y_I(y)$, a (complex)
null vector, $Y^2\equiv Y_I Y_I=0$.  It  carries $U(1)$ weight $(+2)$ and can thus be expressed as  
a polynomial of degree two in the $SU(4)$ harmonic variables $y_{a'}^a$ (see Appendix~\ref{HV}).
The main subject of this paper is the four-point correlation function of the 
operators (\ref{defO}). 

The complete stress-tensor supermultiplet  \p{TT} is obtained by acting on the lowest weight  \p{defO} with half of the $\cN=4$ supersymmetry generators realised in the analytic superspace:
\begin{align}\label{}
\cT(x,\rho,\bar\rho,y) = \exp\left(\rho^a_\a \, Q^\a_a  + \bar\rho^\da_{a'} \, \bar Q^{a'}_\da \right)\cO(x,y)\,.
\end{align}
By construction, this multiplet is annihilated by the other half of the supercharges, hence the name ``half-BPS". In what follows we will drop the dependence on the anti-chiral variables  (for the explanation see Appendix~\ref{sec:an-s_n-invariant}) and will restrict $\cT$ to its purely chiral sector, $\cT(x,\rho,0,y) $. The expansion of this object in the four odd variables $\rho^a_\a$ is very short and its top component is the (on-shell) chiral Lagrangian of $\cN=4$ SYM:
\begin{align}\label{Tch}
\cT(x,\rho,0,y) = \cO(x,y) + \ldots + (\rho)^4 \cL_{\cN=4}(x)\,.
\end{align} 
The explicit form of the complete expansion can be found in Ref.~\cite{paper1}. 
Notice that $\cL_{\cN=4}(x)$ is an $SU(4)$ singlet (hence $y-$independent),  since the entire $U(1)$ weight $(+4)$ of $\cT$  is carried by the odd factor 
$(\rho)^4= \prod_{a,\a} \rho^{\,a}_{\a} $. 
This allows us to write the on-shell action of the $\cN=4$ theory
$S_{\cN=4} = \int d^4x\, \cL_{\cN=4}(x)$
 as a Grassmann integral over the chiral half of analytic superspace   \cite{EHW}
 \begin{align}\label{action}
S_{\cN=4} = \int d^4x \int d^4\rho \ \cT(x,\rho,0,y) \,.
\end{align}
We recall that the action $S_{\cN=4}$ is $y-$independent and it is invariant under the full $\cN=4$ supersymmetry.

\subsection{Correlation functions of the $\cN=4$ stress-tensor multiplet}
\label{sect.2.2}
Let us consider the $n-$point correlation function of the analytic supermultiplets $\cT$ restricted to its chiral sector,  see \p{Tch}. By construction, it depends on $n$ copies
of the (chiral) super-coordinates $(x_i,\rho_i,y_i)$.  Viewed as a function
of the `t Hooft coupling $a=g^2 N_c / (4 \pi^2)$ and Grassmann variables $\rho_i$, the correlation function
admits a double series expansion  
\begin{equation}
G_n=\la \cT(1) \ldots \cT(n) \ra \, = \, \sum_{k=0}^{n-4} \sum_{\ell=0}^\infty a^{\ell+k}
 \, G^{(\ell)}_{n;k}(1,\ldots,n)\,,
\label{corLeftloop}
\end{equation}
where $\cT(i) \equiv  \cT(x_i,\rho_i,0,y_i)$. As was explained in Refs.~\cite{paper1,paper2}, each 
$G_{n;k}^{(\ell)}$ is a homogeneous polynomial in the Grassmann variables
$\rho_i$ (with $i=1,\ldots,n$) of degree ${4k}$. It is accompanied by an additional power  $a^k$ of the coupling, which takes into account the offset of the perturbative level of the Born-type correlation functions with Lagrangian insertions. For $k=0$
the functions $G_{n;0}^{(\ell)}$ define the $\ell-$loop corrections to the correlation
function of the lowest components \p{defO}. The maximal Grassmann
degree allowed by $\cN=4$ superconformal symmetry on the right-hand
side of \p{corLeftloop} is not $4n$, but $4(n-4)$ (see Appendix~\ref{sec:an-s_n-invariant} for details). 

Let us consider the correlation function \p{corLeftloop} at four points, for $n=4$. In this case,
the sum over $k$ reduces to a single 
term with $k=0$, which is $\rho-$independent. In other words, the four-point correlation function of stress-tensor multiplets does not have a purely chiral nilpotent sector. This allows us to replace the superfield  \p{Tch} 
by its lowest component  \p{defO}, so that \p{corLeftloop} becomes simply
\begin{equation}
G_4=\la \cO(x_1,y_1) \ldots  \cO(x_4,y_4)\ra \, = \,  \sum_{\ell=0}^\infty a^{\ell}
 \, G^{(\ell)}_{4}(1,2,3,4)\,.
\label{cor4loop}
\end{equation}
Here $G^{(\ell)}_{4}\equiv G^{(\ell)}_{4;0}$ are the four-point loop corrections, which will be the subject of this paper.

We start with examining \p{cor4loop} at tree level, for $\ell=0$. At this order in the
coupling,
the correlation function \p{cor4loop} reduces to a product of free scalar propagators
\begin{align}
Y_I(y_i) Y_J(y_j) \langle \Phi^I(x_i)  \Phi^J(x_j)\rangle = \frac{(Y(y_i)\cdot  Y(y_j))}{4\pi^2 x_{ij}^2}  =\frac1{4\pi^2}\frac{y_{ij}^2}{x_{ij}^2}\,,
\end{align}
where $y_{ij}^2$ is defined in Appendix~\ref{HV} and $x_{ij}=x_i-x_j$. In this way
we obtain the tree-level expression for the correlation function \p{cor4loop} 
\begin{eqnarray}\label{eq:1}
G^{(0)}_4(1,2,3,4) 
& = & \frac{(N_c^2-1)^2}{4 \, (4 \pi^2)^4}
\left(\frac{y^4_{12}}{x^4_{12}} \frac{y^4_{34}}{x^4_{34}} \, + \,
 \frac{y^4_{13}}{x^4_{13}} \frac{y^4_{24}}{x^4_{24}} \, + \,
 \frac{y^4_{41}}{x^4_{41}} \frac{y^4_{23}}{x^4_{23}} \right) \\
& + & \frac{N_c^2 -1}{(4 \pi^2)^4} \,\
\left( \frac{y_{12}^2}{x_{12}^2} \frac{y_{23}^2}{x_{23}^2}  
\frac{y_{34}^2}{x_{34}^2} \frac{y_{41}^2}{x_{41}^2} \, + \, 
 \frac{y_{12}^2}{x_{12}^2} \frac{y_{24}^2}{x_{24}^2}  
\frac{y_{34}^2}{x_{34}^2} \frac{y_{13}^2}{x_{13}^2} \, + \, 
 \frac{y_{13}^2}{x_{13}^2} \frac{y_{23}^2}{x_{23}^2}  
 \frac{y_{24}^2}{x_{24}^2} \frac{y_{41}^2}{x_{41}^2} \right), \nonumber
\end{eqnarray}
where the first and second lines describe the disconnected and connected
contributions, respectively. 
  
For computing the loop corrections $G^{(\ell)}_{4}$ we employ the
method of Lagrangian insertions. It relies on the observation that the
derivatives of the four-point correlation function  \p{cor4loop} with respect to the coupling constant
can be expressed in terms of the five-point correlation function involving an additional insertions of the $\cN=4$ SYM action, 
\begin{align}
a \frac{\partial}{\partial a} G_4 =  \int d^4 x_5 \, \la \cO(x_1,y_1) \ldots  \cO(x_4,y_4) \cL_{\cN=4}(x_5)
\ra\,.
\end{align}  
The relation is very useful because it allows us to compute the one-loop correction to $G_4$ from the Born-level expression for the five-point correlation function. In the same way, successively differentiating with respect to the coupling, we can express the $\ell-$loop correction to $G_4$ in terms of  the (integrated) Born-level correlation function with $\ell$ insertions of the $\cN=4$ SYM Lagrangian.
Remarkably, since the Lagrangian is itself a member of the stress-tensor multiplet, the latter correlation function can be identified as a particular component
in the expansion of the $(4+\ell)-$point correlation function \p{corLeftloop}
\begin{align}\label{nilG}
G_{4+\ell; \ell}^{(0)}|_{\rho_1=\ldots=\rho_4=0}  = \vev{\cO(x_1,y_1) \ldots \cO(x_4,y_4) \cL(x_5) \ldots \cL(x_{4+\ell})}^{(0)}  (\rho_5)^4 \ldots (\rho_{4+\ell})^4 \,,
\end{align}
where the superscript `$(0)$' indicates the  Born-level approximation.  

In summary, the $\ell-$loop correction to the four-point correlation function \p{cor4loop} 
can be expressed in terms of 
the Born-level $(4+\ell)-$point correlation functions $G_{4+\ell; \ell}^{(0)} $ at Grassmann level ${4\ell}$, 
\begin{equation}\label{eq:3}
G_{4}^{(\ell)}(1,2,3,4) \, = \, \int d^4x_{5} \dots d^4 x_{4+\ell} \left(
\frac{1}{\ell!}  \int d^4\rho_{5} \dots d^4\rho_{4+\ell} \, G_{4+\ell; \ell}^{(0)}(1,\ldots, 4+\ell)\right)\,.
\end{equation}
Here the integration over the Grassmann variables effectively picks out only the component
of $G_{4+\ell; \ell}^{(0)}$ displayed in \p{nilG}. The expression in the parentheses on the right-hand side of \p{eq:3} defines  the {\em integrand} of the loop corrections to the four-point correlation function. 

The purpose of our study will be to determine the {\it integrands
    $G_{4+\ell; \ell}^{(0)}$  without any Feynman graph
    calculations}. Our approach is based on, firstly, fully exploiting
  the symmetries of the integrand. In particular, we reveal a new
  permutation symmetry involving all the $(4+\ell)$ points, which is a
  powerful restriction on  the possible form of the
  integrand. Secondly, using the recently discovered correlation functions/amplitudes  duality~\cite{EKS2,EKS3}, we compare our prediction  for $G_{4+\ell; \ell}^{(0)}$ with the known results on the $\ell-$loop four-gluon scattering amplitude. We demonstrate that this information is sufficient to unambiguously fix the form of the integrand in \p{eq:3}. In this paper we examine in detail the case $\ell=3$ and give some preliminary results on $\ell=4$; the generalisation to arbitrary values of $\ell$ will be considered elsewhere. 

To begin with,  let us recall the known results for the one- and two-loop
integrands~\cite{oneTwo,ESS,Rome} (i.e. the five- and six-point Born-level correlation functions
of Grassmann degree $4$ and $8$, respectively)
written in a particularly suggestive way:  
\begin{align}
  \label{eq:4}
G_{5;1}^{(0)}(1,2,3,4,5)\phantom{,6}& = \frac{2 \, (N^2_c-1)}{(-4 
  \pi^2)^{5}} \times \cI_5 \times{1 \over \prod_{1\leq i<j \leq 5} x_{ij}^2}\,, 
  \\
  G_{6;2}^{(0)}(1,2,3,4,5,6)& = \frac{2 \, (N^2_c-1)}{(-4 
\pi^2)^{6}} \times  \cI_6 \times
{\frac1{48}\sum_{\sigma \in S_6} x_{\sigma_1
      \sigma_2}^2 x_{\sigma_3 \sigma_4}^2x_{\sigma_5 \sigma_6}^2
  \over \prod_{1\leq i<j \leq 6} x_{ij}^2}\,.
\label{eq:5}
\end{align}
Here in the second relation the sum runs over all permutations $(\sigma_1,\ldots,\sigma_6)$ of the indices $(1,2,\ldots,6)$ and the additional factor
 $1/48$  ensures that we do not count the same term more than once.  In relations \p{eq:4} and \p{eq:5}, the dependence on the Grassmann variables is hidden in $\cI_n$  (with $n=5,6$), which is a nilpotent superconformally covariant function of all the superspace points $(x_i, \rho_i, y_i)$. It is described in detail  in Appendix~\ref{sec:an-s_n-invariant}, 
but for the moment we just formulate its three main properties  
needed for our purposes. 

Firstly, $\cI_n$  is a homogeneous polynomial in the Grassmann variables $\rho_1,\ldots,\rho_n$ of degree $4(n-4)$. We will only need the explicit expression for the coefficient in front of
$(\rho_5)^4\ldots(\rho_n)^4$ since this is the only term which contributes to the
superspace integral in (\ref{eq:3}). It is obtained  by setting $\rho_1=\ldots=\rho_4=0$:
\begin{align}\label{eq:6}
 \cI_n|_{\rho_1=\ldots=\rho_4=0} =  (x_{12}^2x_{13}^2 x_{14}^2 x_{23}^2x_{24}^2 x_{34}^2 )\times R(1,2,3,4) \times (\rho_5)^4
 \ldots (\rho_n)^4 \,.
\end{align}
Here $R(1,2,3,4)$ is a rational function of the bosonic (space-time and harmonic) coordinates at the four external points $1,2,3,4$:
 \begin{align}\label{eq:7}
 R(1,2,3,4) &= \frac{y^2_{12}y^2_{23}y^2_{34}y^2_{14}}{x^2_{12}x^2_{23}x^2_{34} x^2_{14}}(x_{13}^2 x_{24}^2-x^2_{12} x^2_{34}-x^2_{14} x^2_{23})\nt
&+\frac{ y^2_{12}y^2_{13}y^2_{24}y^2_{34}}{x^2_{12}x^2_{13}x^2_{24} x^2_{34}}(x^2_{14} x^2_{23}-x^2_{12} x^2_{34}-x_{13}^2 x_{24}^2)  \nt
& +\frac{y^2_{13}y^2_{14}y^2_{23}y^2_{24}}{x^2_{13}x^2_{14}x^2_{23} x^2_{24}}(x^2_{12} x^2_{34}-x^2_{14} x^2_{23}-x_{13}^2 x_{24}^2) \nt &+
\frac{y^4_{12} y^4_{34}}{x^2_{12}x^2_{34}}   + \frac{y^4_{13} y^4_{24}}{x^2_{13}
x^2_{24}} + \frac{y^4_{14} y^4_{23}}{x^2_{14}x^2_{23}} \,.
\end{align}
It is easy to check that $R(1,2,3,4)$ as well as the first factor
on the right-hand side of \p{eq:6} have $S_4$ symmetry under the permutation of the four points. Similarly, the product of odd variables in  \p{eq:6} has
an obvious $S_{n-4}$ symmetry under the exchange of the remaining $(n-4)$ internal (integration) points,
so that the total symmetry of \p{eq:6} is $S_4\times S_{n-4}$. 
 
The second crucial feature of $\cI_n$  is that the complete expression for  it 
as a function of the $n$ superspace points $(x_i, \rho_i, y_i)$ is invariant under 
exchange of any two points, i.e. it  is completely $S_n$ permutation 
symmetric. For the particular component of $\cI_n$ given in \p{eq:6} this symmetry
is broken down to $S_4\times S_{n-4}$, but the $S_n$ symmetry is recovered after restoring the dependence on $\rho_1,\ldots,\rho_4$ (see Appendix~\ref{sec:an-s_n-invariant}).
 
The third feature of $\cI_n$ is that it has (the chiral half of) $\cN=4$  
superconformal symmetry. Namely, it is invariant under chiral super-Poincar\'e 
symmetry and anti-chiral special conformal supersymmetry. In addition, it has conformal weight $(-2)$ and $U(1)$ charge $(+4)$ at  
each point. Most importantly, these properties fix the form of $\cI_n$,  up  to an arbitrary conformally invariant factor depending on the space-time variables $x_i$ only.

Comparing the conformal and $SU(4)$  properties
of $\cI_{4+\ell}$ with those of the correlation function $G_{4+\ell;\ell}^{(0)}$ defined in \p{nilG}, we observe
that they match except for the conformal weight. For the correlation function, the
weight at each point is fixed by the conformal properties of the stress-tensor multiplet \p{Tch}  {to be $(+2)$ whereas for $\cI_{4+\ell}$ it equals $(-2)$}. Therefore, the nilpotent function $\cI_{4+\ell}$ should be
multiplied by a conformally covariant function of $x_i$ with weight $(+4)$ at each 
point. Indeed,  from \p{eq:4} and \p{eq:5} it follows that for $\ell=1$ and $\ell=2$
the correlation function $G_{4+\ell;\ell}^{(0)}$ has the expected form. 

Drawing together these facts, we can
write  $G_{4+\ell;\ell}^{(0)}$ for arbitrary $\ell$ 
in the form  
 \begin{equation}
   \label{eq:8}
    G_{4+\ell;\ell}^{(0)}(1,\dots, 4+\ell) = \frac{2 \, (N^2_c-1)}{(-4
\pi^2)^{4+\ell}} \times  \cI_{4+\ell} \times
 f^{(\ell)}(x_1, \dots, x_{4+\ell})\,,
 \end{equation}
where $f^{(\ell)}$ is a conformally covariant function of $x_1,\ldots,x_{4+\ell}$ with weight $(+4)$ at each
point. {Here the constant factor on the right-hand side of \p{eq:8} is introduced to simplify the expression for 
$f^{(\ell)}$ (see Eq.~\p{f12} below).}
Most importantly, it follows from the full crossing
symmetry of the correlation function \p{corLeftloop} together with the aforementioned
permutation $S_n$ symmetry of the prefactor $\cI_{4+\ell}$ that this function
must be completely symmetric under exchange of any of the $4+\ell$ points:   
\begin{align}\label{permutation}
 f^{(\ell)}(\ldots,x_i, \dots, x_j,\ldots) =  f^{(\ell)}(\ldots,x_j, \dots, x_i,\ldots)  \,.
\end{align}
This is a non-trivial permutation symmetry, specific to the $\cN=4$ theory.\footnote{Explicit perturbative calculations \cite{ESS} in a generic conformal $\cN=2$ theory show that in general the two-loop integrand does not have this symmetry, it is only there in the $\cN=4$ case.}
We will show below that the $S_{4+\ell}$ symmetry, together with the conformal and some other properties, impose strong 
constraints on the possible form of the function $f^{(\ell)}$. Our aim
is to restrict it to a unique expression using the correlation
function/amplitude duality of \cite{EKS2,EKS3}. In this paper we do this in the cases $\ell=3,4$.

\begin{figure}[ht]
\psfrag{s1}[cc][cc]{${}_1$}
\psfrag{s2}[cc][cc]{${}_2$}
\psfrag{s3}[cc][cc]{${}_3$}
\psfrag{s4}[cc][cc]{${}_4$}
\psfrag{s5}[cc][cc]{${}_5$}
\psfrag{i1}[cc][cc]{$\sigma_1$}
\psfrag{i2}[cc][cc]{$\sigma_2$}
\psfrag{i3}[cc][cc]{$\sigma_3$}
\psfrag{i4}[cc][cc]{$\sigma_4$}
\psfrag{i5}[cc][cc]{$\sigma_5$}
\psfrag{i6}[cc][cc]{$\sigma_6$}
\psfrag{a}[cc][cc]{$\ell=1$}
\psfrag{b}[cc][cc]{$\ell=2$}
\centerline{\includegraphics[height=45mm]{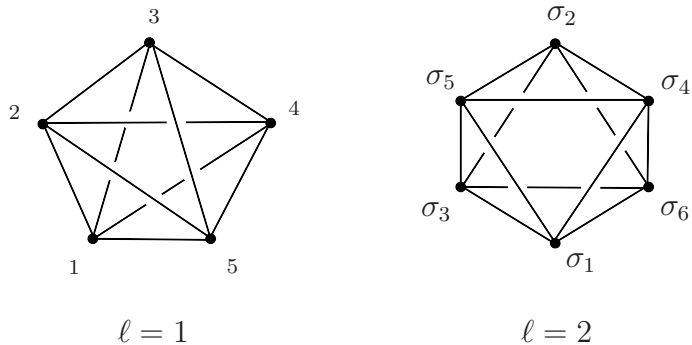}}
\caption{\small Diagrammatic representation of the function $f^{(\ell)}(x)$ for $\ell=1$ and $\ell=2$,  Eq.~(\ref{f12}). Each line with labels $i$ and $j$ at the end points denotes a scalar
propagator $1/x_{ij}^2$.}
\label{1+2-loop}
\end{figure}

For $\ell=1$ and $\ell=2$ we find from \p{eq:4} and \p{eq:5} the
explicit expressions for the function 
\begin{align}\notag
 f^{(1)}(x_1,\ldots,x_5) &= {1 \over \prod_{1\leq i<j \leq 5}   x_{ij}^2} \,,
\\ \label{f12}
 f^{(2)}(x_1,\ldots,x_6) &= \frac1{48}\sum_{\sigma \in S_6} {x_{\sigma_1
      \sigma_2}^2 x_{\sigma_3 \sigma_4}^2x_{\sigma_5 \sigma_6}^2\over \prod_{1\leq i<j \leq 6}   x_{ij}^2}  \,.
\end{align}
These relations can be represented in a diagrammatic form as 
shown in Fig.~\ref{1+2-loop}.

We recall that the Borm-level correlation function $G_{4+\ell;\ell}^{(0)}$ defines
the integrand of the four-point correlation function \p{eq:3} at $\ell$ loops. Substituting
\p{eq:8} into \p{eq:3} we find that the  $\ell-$loop correction to the  four-point correlation function takes the form 
\begin{align}\label{intriLoops}
  G_{4}^{(\ell)}(1,2,3,4)= \frac{2 \, (N_c^2-1)}{(4\pi^2)^{4}} \times R(1,2,3,4)   \times  F^{(\ell)} \qquad \mbox{for $\ell \ge 1$} \,,
\end{align}
where $R(1,2,3,4)$ is given by \p{eq:7} and the function $F^{(\ell)}$ is defined by the Euclidean integral 
\begin{align}\label{integg}
  F^{(\ell)}(x_1,x_2,x_3,x_4)={x_{12}^2 x_{13}^2 x_{14}^2 x_{23}^2 
  x_{24}^2 x_{34}^2\over \ell!\,(-4\pi^2)^\ell} \int d^4x_5 \dots
  d^4x_{4+\ell} \,  f^{(\ell)}(x_1,  \dots , x_{4+\ell})\ .
\end{align}
Taking
into account the conformal properties of the function $f^{(\ell)}$, we find from \p{integg} that $F^{(\ell)}$ is a conformally covariant function of $x_1,x_2,x_3,x_4$ with
weight $(+1)$ at each point. The prefactor $R(1,2,3,4)$ has the same conformal weights. Thus, their product on the right-hand side of \p{intriLoops} has conformal weights $(+2)$ as it should be for the four-point correlation function $G_4(1,2,3,4)$.

The presence of the universal rational prefactor $R$ in \p{intriLoops}  for any $\ell$ signifies that the loop corrections to the four-point correlation function of $\cN=4$ stress-tensor multiplets is determined by a single function of the four points $x_1,\ldots,x_4$. This is much less freedom than what one could expect in the six $SU(4)$ channels of the tensor product $\mathbf{20'}\times\mathbf{20'} = \mathbf{1} + \mathbf{15} + \mathbf{20'} + \mathbf{84} + \mathbf{105} +\mathbf{175}$. This fact is known under the name of ``partial non-renormalisation theorem" \cite{partialNonRen}. 

The relation \p{integg} illustrates the non-trivial meaning of the permutation symmetry \p{permutation} of the function $f^{(\ell)}$. In application to \p{integg}, it
exchanges the external points $x_1,x_2,x_3,x_4$ and the integration points $x_5, \dots, x_{4+\ell}$ and, therefore, puts the two sets of points on an equal footing.
We would like to emphasise that this symmetry is obscured {for the function $F^{(\ell)}$ } by the integration
on the right-hand side of \p{integg}, since it breaks
the $S_{4+\ell}$ symmetry of the function $f^{(\ell)}$ down to the $S_4$
symmetry of the function $F^{(\ell)}$.
 
\section{The four-point correlation function at three loops}
\label{sec:three-loop-4}

In this section, we combine the permutation symmetry of the integrand 
together with the conjectured correlation function/amplitude
duality to work out the three-loop expression for the four-point correlation
function \p{cor4loop}.

\subsection{General Ansatz for the three-loop integrand}

Let us find out to what extent we can determine the rational  function
$f^{(\ell)}$ in \p{integg}, for example at three loops, for $\ell=3$. To do this 
we examine again the one- and two-loop functions given in (\ref{f12}) and 
note that they have a very specific denominator of the form $\prod_{i<j} x_{ij}^2$. 
As we will see in a moment, this form of the denominator can be deduced
from the operator product expansion (OPE) analysis of the correlation function
\p{eq:8} and it should be universal for any $\ell$. This suggests to write the general 
Ansatz for the function $f^{(\ell)}$ as follows:
\begin{equation}
   \label{eq:10}
   f^{(\ell)}(x_1, \dots, x_{4+\ell})= { P^{(\ell)}(x_1, \dots ,
   x_{4+\ell}) \over \prod_{1\leq i<j \leq 4+\ell}   x_{ij}^2}\ ,
 \end{equation} 
where $P^{(\ell)}$ is a homogeneous polynomial in $x_{ij}^2$ which is invariant under $S_{4+\ell}$ permutations of $x_i$. We recall that the
function $f^{(\ell)}$ is conformally covariant with weight $(+4)$ at each point, while the denominator on the right-hand side of \p{eq:10} has weight $-(\ell+3)$. As a consequence, the polynomial $P^{(\ell)}$ should
have uniform weight $-(\ell-1)$ at each point, both external and internal. This
implies, in particular, that $P^{(\ell)}$ is a polynomial in $x_{ij}^2$ of degree $(\ell-1)(\ell+4)/2$. Indeed, the one- and two-loop expressions \p{f12} have this property. At three loops, for $\ell=3$, the degree of $P^{(3)}$ should equal $7$. 

{A characteristic feature of the Ansatz (\ref{eq:10}) is that the distances between each pair of points appear in the denominator on the right-hand side of  (\ref{eq:10})  to the first
power only. In other words, for a generic polynomial $P^{(\ell)}$ the function
 (\ref{eq:10}) scales as $f^{(\ell)}(x)\sim 1/x_{ij}^2$ for $x_i\to x_j$. This 
 property can be understood as follows. We recall that according to \p{nilG}, \p{eq:6} and (\ref{eq:8})
the function $f^{(\ell)}(x)$ determines the {\em Born-level} expression for 
the correlation function involving  the protected  operator $\mathcal{O}$ and the  $\cN=4$ SYM Lagrangian $\mathcal{L}$ (itself protected, since it is a member of the same supermultiplet as $\cO$):
\begin{align}\label{OPE-ans}
\vev{\mathcal{O}(1) \ldots  \mathcal{O}(4)\mathcal{L}(5)\ldots \mathcal{L}(4+\ell) }^{(0)} 
\sim  R(1,2,3,4) \,
x_{12}^2 x_{13}^2 x_{14}^2 x_{23}^2 x_{24}^2 x_{34}^2\,
 f^{(\ell)}(x)  \,.
\end{align}
As a result, the asymptotic behaviour of the function $f^{(\ell)}(x)$ for $x_i \to x_j$ 
should follow from the {(Born-level)} OPE expansion of the operators located at the points $x_i$ and $x_j$. Depending on the choice of points, we can distinguish three different cases:
$\mathcal{O}(i)\mathcal{O}(j)$, $\mathcal{O}(i)\mathcal{L}(j)$ and 
$\mathcal{L}(i)\mathcal{L}(j)$. 

In the first case, the OPE expansion of the product of two half-BPS operators $\mathcal{O}$  is specified in Eq.~(\ref{ope}) below. For $x_i\to x_j$ it is
dominated by the contribution from the identity operator $\mathcal{O}(1)\mathcal{O}(2)\sim \cI/x_{12}^4+O(1/x_{12}^2)$. However, the identity operator only 
contributes to the disconnected part of the correlation function, $\vev{\mathcal{O}(1)\mathcal{O}(2)}\vev{\mathcal{O}(3)   \mathcal{O}(4)\mathcal{L}(5)\ldots \mathcal{L}(4+\ell) }$. As a consequence, the leading contribution to the connected correlation function in (\ref{OPE-ans}) comes from subleading $O(1/x_{12}^2)$ terms in the OPE (see Eq.~(\ref{ope})), thus implying that the right-hand
side of (\ref{OPE-ans}) should scale as $1/x_{12}^2$. Indeed, it follows from (\ref{eq:7})
that $R(1,2,3,4)\sim  1/x_{12}^2$  and, therefore, $x_{12}^2 f^{(\ell)}(x)$ should stay finite for $x_1\to x_2$. This leads to $f^{(\ell)}(x)\sim 1/x_{12}^2$ in agreement with (\ref{eq:10}). Let us now examine the OPE
expansion of $\mathcal{O}(1)\mathcal{L}(j)$.  For $x_1\to x_j$ the leading contribution
scales as $c\, \mathcal{O}(1)/x_{1j}^4$. However, the corresponding coefficient function is proportional to the three-point function $c \sim \vev{\mathcal{O}\mathcal{O}\mathcal{L}}$ which vanishes in $\mathcal{N}=4$ SYM to all loops (non-renormalisation of the protected two-point function $\vev{\cO \cO}$). As a result, the dominant contribution to the OPE scales as $1/x_{1j}^2$, again leading to the asymptotic behaviour  $f^{(\ell)}(x)\sim 1/x_{1j}^2$.
Finally, we consider the OPE expansion of the product of two Lagrangians. It is known that the dominant contribution to $\mathcal{L}(i)\mathcal{L}(j)$ as $x_i\to x_j$ comes from
contact terms proportional to $\delta^{(4)}(x_i-x_j)$ and its derivatives. Since the 
correlation function (\ref{OPE-ans}) involves Lagrangians at distinct points, the contact
terms do not contribute to  (\ref{OPE-ans}) and we are left with the $1/x_{ij}^2$ contribution only.%
\footnote{The contact terms play an important role in the Lagrangian insertion procedure 
 based on the differentiation of the correlation function with respect to the coupling
 constant. Namely, they ensure that successive differentiation just inserts the action without reproducing the correlation function itself. }
This immediately implies that $f^{(\ell)}(x)\sim 1/x_{ij}^2$. {The above considerations explain the form of the denominator in our Ansatz \p{eq:10}.}
 } 

\begin{figure}[h!]
\vspace*{5mm}
\psfrag{i1}[cc][cc]{$\sigma_1$}
\psfrag{i2}[cc][cc]{$\sigma_2$}
\psfrag{i3}[cc][cc]{$\sigma_3$}
\psfrag{i4}[cc][cc]{$\sigma_4$}
\psfrag{i5}[cc][cc]{$\sigma_5$}
\psfrag{i6}[cc][cc]{$\sigma_6$}
\psfrag{i7}[cc][cc]{$\sigma_7$}
\psfrag{a}[cc][cc]{$\rm (a)$}
\psfrag{b}[cc][cc]{$\rm (b)$}
\psfrag{c}[cc][cc]{$\rm (c)$}
\psfrag{d}[cc][cc]{$\rm (d)$}
\centerline{\includegraphics[height=47mm]{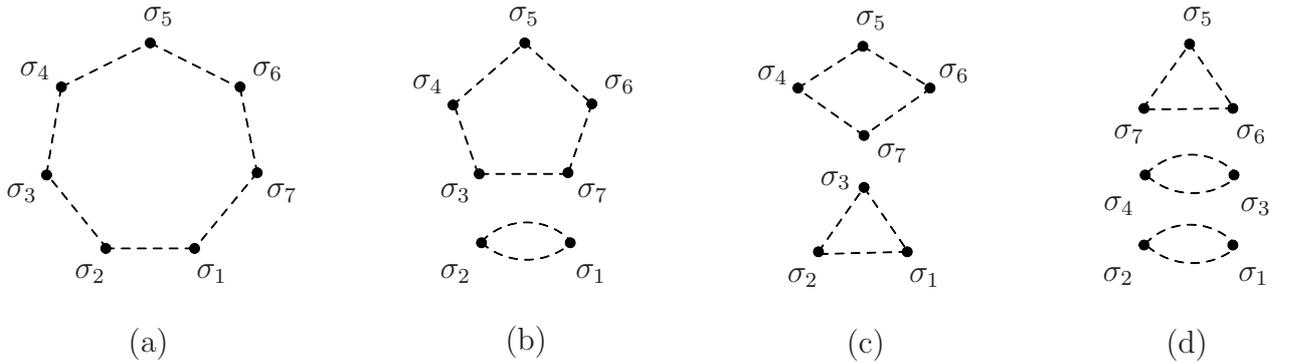}}
\caption{\small  Diagrammatic representation of the different $S_7$ symmetric 
polynomials  $P^{(3)}(x_i)$ defined in (\ref{eq:11}). 
The indices $(\sigma_1,\ldots,\sigma_7)$ correspond to all permutations
of the external points $(1,\ldots,7)$.
The dashed lines with indices $\sigma_i$ and $\sigma_j$ at the end  points denote the factors $x_{\sigma_i\sigma_j}^2$. }
\label{fig:1} 
\end{figure}

Making use of the Ansatz \p{eq:10} for $\ell=3$, we can reduce the problem
of constructing the rational function $f^{(3)}(x_1, \dots, x_7)$  to that of finding
$S_7$ symmetric conformally covariant polynomial $P^{(3)}(x_1 \dots, x_7)$ of
weight $(-2)$. As was already mentioned, $P^{(3)}$ takes the form of a
homogeneous polynomial of $x_{ij}^2$ of degree $7$ and it is not hard
to convince oneself that there are in fact only four 
possibilities~\footnote{In fact, the polynomials listed in \p{eq:11} are not independent in four dimensions due to a single Gram determinant constraint relating all four of them \cite{Eden:2012tu}.} \footnote{In principle, with six points or more one
  can form pseudo-scalar conformal invariants which could appear in
  the integrand. However, these involve $\epsilon$ tensors, which are
  antisymmetric under interchanges and are thus
  inconsistent with the aforementioned $S_{4+\ell}$ permutation symmetry. Moreover,  the four-point correlation function (the integral \p{integg} itself) cannot contain parity-odd terms. This shows that, even if such terms exist at the level of the integrand, they must be total derivatives which do not contribute to the integral. For this reason we do not consider pseudo-scalars in \p{eq:11}. }  for $P^{(3)}$: 
\begin{align}
 & {\text{(a) heptagon:}} &&x_{12}^2 x_{23}^2 x_{34}^2
  x_{45}^2 x_{56}^2 x_{67}^2  x_{71}^2  \ +\ {S_7\ \mathrm{permutations}}\,, \nonumber
 \\[2mm]
 & {\text{(b) 2-gon  $\times$ pentagon:}} &&(x_{12}^4)( x_{34}^2 x_{45}^2 x_{56}^2 x_{67}^2 x_{73}^2)  \ +\ {S_7\ \mathrm{permutations}} \,, 
 \nonumber
 \\[2mm]
   & {\text{(c) triangle $\times$  square:}} &&(x_{12}^2  x_{23}^2 x_{31}^2) ( x_{45}^2
  x_{56}^2 x_{67}^2 x_{74}^2 ) \ +\ {S_7\ \mathrm{permutations}}   \,,
  \nonumber
  \\[2mm]
 & {\text{(d) 2-gon $\times$ 2-gon $\times$ triangle:}} &&(x_{12}^4)( x_{34}^4)( 
   x_{56}^2 x_{67}^2 x_{75 }^2)  \ +\ {S_7\ \mathrm{permutations}}\, .\label{eq:11}
\end{align}
It is convenient to represent the different choices in the form of diagrams as shown
in Fig.~\ref{fig:1}.
 
\begin{figure}[h]
\psfrag{i1}[cc][cc]{$\sigma_1$}
\psfrag{i2}[cc][cc]{$\sigma_2$}
\psfrag{i3}[cc][cc]{$\sigma_3$}
\psfrag{i4}[cc][cc]{$\sigma_4$}
\psfrag{i5}[cc][cc]{$\sigma_5$}
\psfrag{i6}[cc][cc]{$\sigma_6$}
\psfrag{i7}[cc][cc]{$\sigma_7$}
\psfrag{a}[cc][cc]{$\rm (a)$}
\psfrag{b}[cc][cc]{$\rm (b)$}
\psfrag{c}[cc][cc]{$\rm (c)$}
\psfrag{d}[cc][cc]{$\rm (d)$}
\centerline{\includegraphics[height=45mm]{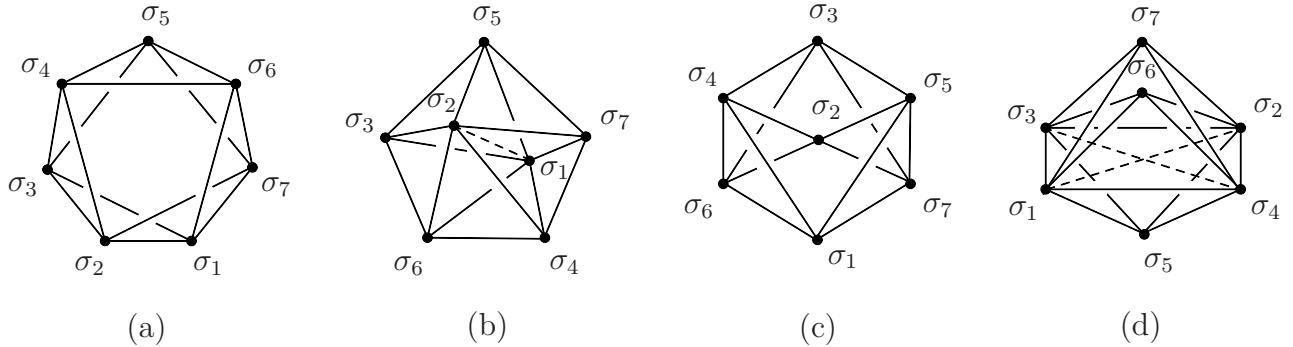}}
\caption{\small  
Diagrammatic representation of the four classes of functions $f^{(3)}(x)$ corresponding to the polynomials shown in Fig.~\ref{fig:1}.  Solid
lines denote scalar propagators $1/x_{\sigma_i\sigma_j}^2$ while dashed
lines stand for  numerator factors $x_{\sigma_i\sigma_j}^2$.}  
\label{fig2}
\end{figure}

In the expression for the function $f^{(3)}$, Eq.~\p{eq:10}, the polynomial 
$P^{(3)}$ is divided by the product of distances between any pair of points.
Replacing $P^{(3)}$ in \p{eq:10} by the various expressions defined  in \p{eq:11},
we find that some factors $x^2_{ij}$ in the numerator cancel out against similar factors in the denominator in  \p{eq:10}. In diagrammatic terms, a solid line and a dashed line between two points annihilate each other. 
The resulting four different contributions to $f^{(3)}$ are represented graphically in Fig.~\ref{fig2}.  An important
feature of these diagrams is that for each vertex the number of solid lines minus the number of dashed lines attached to it equals $4$. This property ensures that 
the conformal weight of the corresponding contribution to $f^{(3)}$ equals $4$  
 at each point.   
 
The general expression for the polynomial $P^{(3)}$ satisfying the symmetry
constraints specified above is given by a linear combination of the four terms  in \p{eq:11} with arbitrary coefficients. Similarly, the general expression for $f^{(3)}$ is given by the same linear combination of four diagrams shown in Fig.~\ref{fig2}. 
{We would like to emphasise that this result holds for a gauge group $SU(N_c)$ with arbitrary $N_c$. }

To fix the value of these coefficients we have to impose some additional conditions. One possible approach is to try to extract some anomalous dimensions, for instance that of the Konishi operator, from the OPE analysis of the correlation function. This requires the ability to compute the integrals whose integrands are encoded in Fig.~\ref{fig2} in the singular {  short-distance limit $x_i\to x_j$}. We explore this approach in Section~\ref{sec:ope-test} in the simpler case of the {\it planar} correlation function (i.e., for large $N_c$), although in principle one could generalise it to the non-planar case as well. 

Another approach consists in exploiting the conjectured duality between scattering amplitudes and correlation 
functions \cite{EKS2,EKS3}.  Since this duality only works for planar amplitudes and correlation functions, we are thus restricting our study to the planar sector. As we show in the next subsection, the comparison of the correlation function with the known result for the three-loop four-point amplitude allows us to unambiguously fix the coefficients of all four topologies in Fig.~\ref{fig2}.  
\subsection{Correlation function/scattering amplitude duality} 
 
In application to the four-point correlation function, the duality establishes the correspondence between the four-particle scattering amplitude $A_4$ in planar $\cN=4$ SYM and the correlation function
$G_4$ in the limit where the four operators become light-like separated in a sequential
fashion, $x_{12}^2=x_{23}^2=x_{34}^2=x_{41}^2=0$:
\begin{align}\label{du}
\lim_{x^2_{i,i+1} \to 0} (G_4(x)/G^{(0)}_4(x)) = (A_4(p)/A^{(0)}_4(p))^2\,.
\end{align} 
Here on the right-hand side the amplitude $A_4(p_1,p_2,p_3,p_4)$ depends on the  light-like momenta of the scattered particles, $p_i^2=0$. They are identified with the coordinates $x_i$ of the operators on the left-hand side through the dual space relation   $p_i=x_i-x_{i+1}$ with the periodicity condition $x_{i+4}=x_i$. 
This duality is understood at the level of the {\it integrands} on both sides of the relation, and not in terms of the divergent {\it integrals}. 

We recall that the $\ell-$loop correction to the correlation function $G_4$ is 
determined by the scalar function  $F^{(\ell)}(x_1,x_2,x_3,x_4)$ defined in 
Eqs.~\p{intriLoops} and \p{integg}. Similarly, the $\ell-$loop correction
to the four-particle amplitude in the planar $\cN=4$ SYM theory  is given by the scalar function $M^{(\ell)}(p_1,p_2,p_3,p_4)= A^{(\ell)}_4(p)/A^{(0)}_4(p)$. At present, the functions $M^{(\ell)}$ are known up to five loops,  for $\ell=1,\ldots,5$, in the form of a sum of scalar $\ell-$loop planar Feynman integrals of various topologies  \cite{Anastasiou:2003kj,bern42,4loopMHV,5loops}.  The duality \p{du} can then be formulated as a relation between the {\em integrands} in $F^{(\ell)}$ and $M^{(\ell)}$ as follows. In the light-like limit we have,
from~(\ref{eq:7}) and~(\ref{eq:1}),
\begin{align}
 R(1,2,3,4) \rightarrow 
x_{13}^2 x_{24}^2\ \frac{y^2_{12}y^2_{23}y^2_{34}y^2_{14}}{x^2_{12}x^2_{23}x^2_{34}
   x^2_{14}}\,, \qquad 
G_4^{(0)}(1,2,3,4)  \rightarrow \frac{N_c^2 -1}{(4 \pi^2)^4} \
 \frac{y^2_{12}y^2_{23}y^2_{34}y^2_{14}}{x^2_{12}x^2_{23}x^2_{34}
   x^2_{14}}\,,  
\end{align}
hence
\begin{align}\label{}
\lim_{x^2_{i,i+1} \to 0} (G_4^{(\ell)}/G^{(0)}_4) = \lim_{x^2_{i,i+1} \to 0} (2x_{13}^2 x_{24}^2 F^{(\ell)})\,.
\end{align} 
Then the duality relation \p{du} becomes
\begin{align}\label{cor-amp}
\lim_{x_{i,i+1}^2\to 0 } \left(1 + 2 x_{13}^2 x_{24}^2  \sum_{\ell\ge 1}  a^\ell F^{(\ell)}  \right) =  
\left(1+\sum_{\ell\ge 1}  a^\ell M^{(\ell)}\right)^2\,,
\end{align}
or equivalently 
\begin{align}\label{spec1}
\lim_{x_{i,i+1}^2\to 0 } (x_{13}^2 x_{24}^2 F^{(1)}) & =  M^{(1)}\,,
\\ \label{spec2}
\lim_{x_{i,i+1}^2\to 0 } (x_{13}^2 x_{24}^2 F^{(2)}) & =  M^{(2)} + \frac12 \left(M^{(1)}\right)^2\,,
\\[2mm] \label{spec}
\lim_{x_{i,i+1}^2\to 0 } (x_{13}^2 x_{24}^2 F^{(3)}) & = M^{(3)} + M^{(1)} M^{(2)}\,, \ \ldots \ \,.
\end{align}  
The first two relations, Eqs.~\p{spec1} and \p{spec2},  have been verified in Ref.~\cite{EKS2} by making use of the explicit two-loop expressions for the correlation function and scattering amplitude. Our task now is to do the next step and use the new relation \p{spec} to fix the coefficients of the four topologies in Fig.~\ref{fig2}.

For the three-loop correlation function, we deduce from \p{integg} that the integrand of $F^{(3)}$ is given by 
\begin{align}\label{F3-int} 
[F^{(3)}]_{\rm integrand} = {x_{12}^2 x_{13}^2 x_{14}^2 x_{23}^2 
  x_{24}^2 x_{34}^2\over 3!\,(-4\pi^2)^3}  \times  f^{(3)}(x_1,  \dots , x_{7})\,,
\end{align}
with $f^{(3)}$ given by a linear combination of graphs shown in Fig.~\ref{fig2}. In the light-like limit, $x_{i,i+1}^2\to 0$, the prefactor on the right-hand side
of \p{F3-int} vanishes but (some of the terms in) the function $f^{(3)}$ develops poles in $1/x_{i,i+1}^2$,
so that their product stays finite. The surviving terms will appear on the right-hand side   in Eq.~\p{spec}.

\begin{figure}[ht]
\vspace*{5mm}
\psfrag{a}[cc][cc]{(a)}
\psfrag{b}[cc][cc]{(b)}
\psfrag{c}[cc][cc]{(c)}
\psfrag{d}[cc][cc]{(d)}
\centerline{\includegraphics[height=45mm]{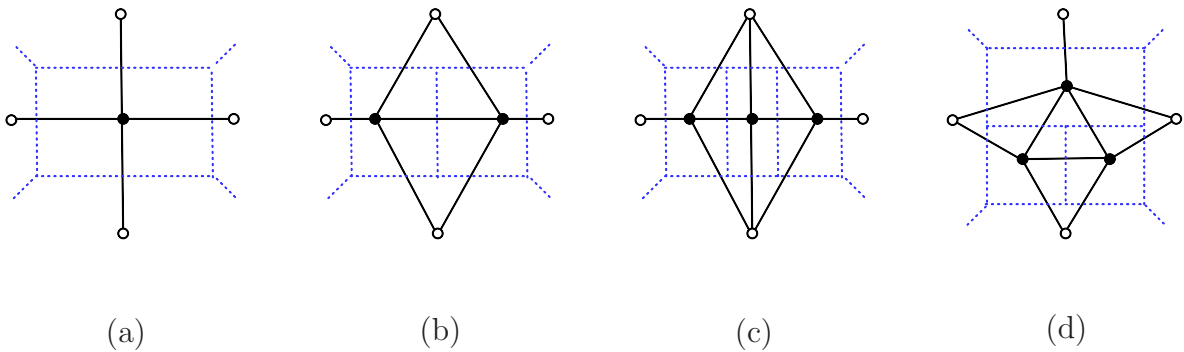}}
\caption{\small  
Dual conformal $x-$integrals (solid lines) and momentum $p-$integrals (doted lines) 
in the three-loop planar four-gluon amplitude:
(a) one-loop ladder, (b) two-loop ladder, (c) three-loop ladder and (d) tennis
court. 
}
\label{fig:3ampl}
\end{figure}

{
Let us find out which integrals we expect to see in  \p{spec}. 
The one- and two-loop four-gluon amplitudes, $M^{(1)}$ and $M^{(2)}$, 
involve one- and two-loop ladder diagrams shown in Fig.~\ref{fig:3ampl} (a)
and (b), respectively. The three-loop four-gluon amplitude $M^{(3)}$ involves only two integral topologies, the three-loop ladder and the so-called  ``tennis court"   \cite{bern42}. They are depicted  in Fig.~\ref{fig:3ampl} (c) and (d), respectively, both as conventional momentum $p-$space diagrams and as dual $x-$space diagrams. }

\begin{figure}[h]
\vspace*{5mm}
\psfrag{a}[cc][cc]{$g(1,2,3,4)$}
\psfrag{b}[cc][cc]{$h(1,3;2,4)$}
\centerline{\includegraphics[height=38mm]{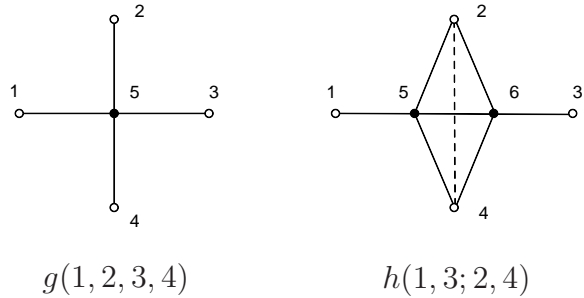}}
\caption{\small  Diagrammatic representation of the one- and two-loop integrals (\ref{eq:g+h}). The solid lines denote scalar propagators and
the dashed line denotes a factor of $x^2_{24}$ in the numerator. Each external point has conformal weight $(+1)$. }
\label{fig:g+h}
\end{figure}

\begin{figure}[ht]
\psfrag{a}[cc][cc]{$T(1,3;2,4)$}
\psfrag{b}[cc][cc]{$E(1,3;2,4)$}
\psfrag{c}[cc][cc]{$L(1,3;2,4)$}
\psfrag{d}[cc][cc]{$g \times h (1,3;2,4)$}
\psfrag{e}[cc][cc]{$H(1,2;3,4)$}
\centerline{\includegraphics[height=40mm]{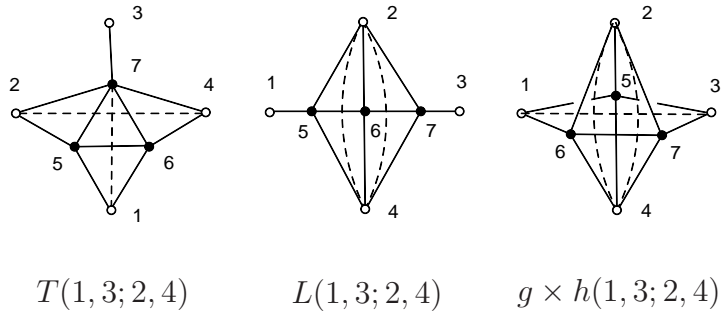}}
\caption{\small  
Diagrammatic representation for the three-loop integrals (\ref{eq:14}) surviving in the light-like limit. The dashed lines (numerator factors) ensure uniform conformal weight $(+1)$ at each external point, as well as at the integration point 7 for $T$. 
}
\label{fig:3onshell}
\end{figure}

The one- and two-loop graphs are redrawn  again in Fig.~\ref{fig:g+h} with the external and internal points labelled and, in the case of the two-loop ladder, with an extra numerator factor (dashed line) added.  This factor balances the conformal weights at points 2 and 4, so that the integral has uniform conformal weight  $(+1)$ at each external point. Similarly, the tennis court $T$ and three-loop ladder $L$ have been redrawn in Fig.~\ref{fig:3onshell} with the necessary dashed lines added. Fig.~\ref{fig:3onshell} contains another diagram of three-loop topology, the product $g\times h$ of the one- and two-loop ladders that we expect to find in the non-linear term $M^{(1)} M^{(2)}$ on the right-hand side of \p{spec}.

We may interpret the graphs in Fig.~\ref{fig:3onshell}, with conformal weight $(+1)$ at each external point,  as those terms in the function $F^{(3)}$, Eq.~\p{F3-int}, which survive in the light-like limit. Now, we are interested in the manifestly $S_7$ symmetric function $f^{(3)}$. In order to upgrade the $F-$terms shown in Fig.~\ref{fig:3onshell} to $f-$terms, we need to divide them by the prefactor  $x_{12}^2 x_{13}^2 x_{14}^2 x_{23}^2 x_{24}^2 x_{34}^2$ from \p{F3-int}. In graphical terms, this means to superpose a square with diagonals (all lines solid) and with vertices at points $1,2,3,4$,  onto each graph in Fig.~\ref{fig:3onshell}. In the process some numerator factors (dashed lines) cancel against propagators (solid lines). The result of this manipulation are graphs of the type shown in Fig.~\ref{fig2}. It is not difficult to see that all three $F-$graphs from Fig.~\ref{fig:3onshell}, when upgraded to  $f-$graphs, fall into the topology \ref{fig2}(b).

We conclude that the three other topologies shown 
 in Fig.~\ref{fig2}(a),(c),(d) should not appear in the light-like limit, otherwise they would give unwanted contributions to the amplitude. It is not hard to check that none of the topologies  in Fig.~\ref{fig2} entirely vanishes in the light-like limit. Hence, the amplitude/correlation function duality leads us to setting the coefficients of the topologies in Fig.~\ref{fig2}(a),(c),(d) to zero.\footnote{We repeat that  this only concerns the correlation function in the planar limit, to which the duality applies. In the general case (for an arbitrary gauge group $SU(N_c)$) all four topologies   in Fig.~\ref{fig2} are expected to contribute to the correlation function.} The only coefficient which still needs to be fixed is that of  the topology  in Fig.~\ref{fig2}(b).  We do this in Section~\ref{square} by examining the details of the duality relation \p{spec}.

{We wish to point out that there exists an alternative way to interpret the duality relation \p{cor-amp}. We may say that the entire planar 
amplitude or correlation function is  generated, up to three loops,  by the lower-loop non-linear terms $(M^{(1)})^2$ in \p{spec2} and $M^{(1)} M^{(2)}$ in \p{spec}. Indeed, at one loop the only possible choice for the numerator of $f^{(1)}$  in \p{f12} is a constant whose value is fixed by the relation \p{spec1}. Then, at two loops the non-linear term $(M^{(1)})^2$ in \p{spec2} appears, together with $M^{(2)}$ (with exactly these coefficients), in the unique symmetric choice for $f^{(2)}$ in \p{f12}. This means that the non-linear term actually tells us what the linear two-loop term  $M^{(2)}$ must be, and hence determines the two-loop amplitude. Similarly, the non-linear term $M^{(1)} M^{(2)}$ in \p{spec} determines the three-loop term $M^{(3)}$, by the simple fact that they belong to the unique symmetric topology 3(b) contributing to $f^{(3)}$.  In Sect.~\ref{Conc} we comment further  on the extent to which higher-loop amplitudes/correlation functions can be determined from lower-loop amplitudes in this way more generally.  }

\subsection{Three-loop integrand}  
   
As was explained in the previous subsection, the three-loop integrand for the planar four-point
correlation function can only receive contributions from the topology shown in Fig.~\ref{fig2}(b). The corresponding expression for the function $f^{(3)}$ reads
(up to an overall normalisation factor)
 \begin{align}\label{f3}
f^{(3)}(x_1,\dots,x_7)&={{1 \over 20} \sum_{\sigma \in S_7} x_{\sigma_1\sigma_2}^4 x_{\sigma_3\sigma_4}^2 x_{\sigma_4\sigma_5}^2 x_{\sigma_5\sigma_6}^2 x_{\sigma_6\sigma_7}^2
   x_{\sigma_7\sigma_3}^2 \over \prod_{1\leq i<j \leq 7}
    x_{ij}^2} \,,
\end{align}  
where the factor $1/20$ simply ensures a sum over distinct terms: there are $7!/20 = 252$ distinct terms in the sum over $S_7$ permutations.

\begin{figure}[h!]
\vspace*{5mm}
\psfrag{a}[cc][cc]{$\rm (a)$}
\psfrag{b}[cc][cc]{$\rm (b)$}
\psfrag{c}[cc][cc]{$\rm (c)$}
\psfrag{d}[cc][cc]{$\rm (d)$}
\psfrag{e}[cc][cc]{$\rm (e)$}
\centerline{\includegraphics[height=40mm]{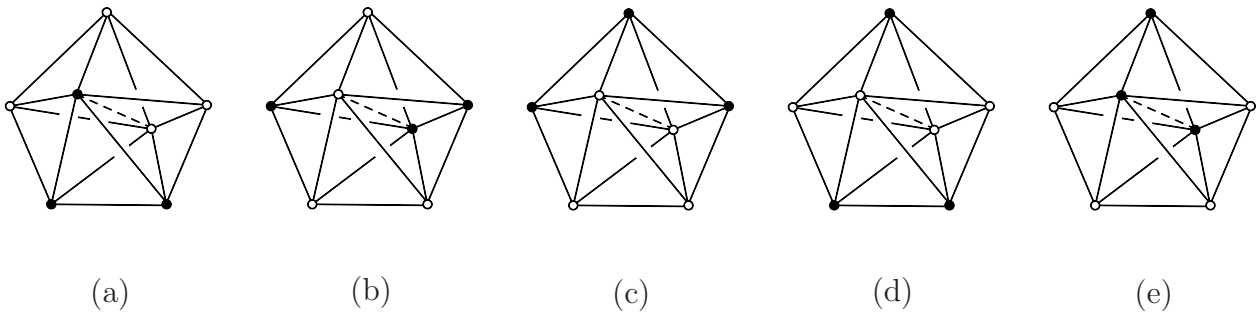}}
\caption{\small  
Diagrams contributing to the function $f^{(3)}$ in the three-loop integrand. White nodes represent the external points $x_1,\dots,x_4$
and black nodes represent the internal points
 $x_5,x_6,x_7$. }
\label{fig-den}
\end{figure}

Defining the three-loop integrand \p{F3-int} we have to split the seven points $x_i$ into 
four external points $x_1,\ldots,x_4$ and three internal (integration) points $x_5,x_6,x_7$.
Then, the $S_7$ symmetric rational function $f^{(3)}$ can be decomposed
into a sum of terms separately invariant under $S_4\times S_3$. Here $S_4$ and $S_3$
act on the external and internal points, respectively. This gives the five distinct terms
displayed in Figure~\ref{fig-den}.
Notice  that, drawn like this, all the graphs  in Fig.~\ref{fig-den} have the same topology, the
 only difference being the localisation of the external (white) and
 internal (black) points. 
 
\begin{figure}[ht]
\psfrag{a}[cc][cc]{$T(1,3;2,4)$}
\psfrag{b}[cc][cc]{$E(1,3;2,4)$}
\psfrag{c}[cc][cc]{$L(1,3;2,4)$}
\psfrag{d}[cc][cc]{$g \times h (1,3;2,4)$}
\psfrag{e}[cc][cc]{$H(1,2;3,4)$}
\centerline{\includegraphics[height=40mm]{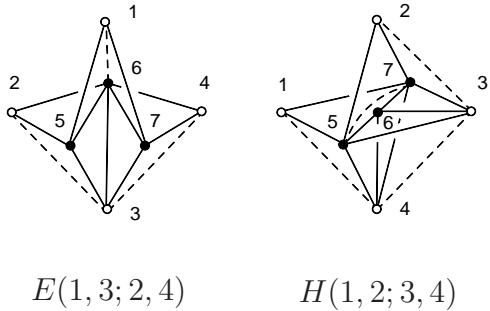}}
\caption{\small  
Diagrammatic representation for the new three-loop integrals (\ref{eq:14}) contributing to the four-point correlation function away from the light-like limit. The dashed lines between adjacent external points indicate that these integrals vanish in the limit. 
}
\label{fig:3offshell}
\end{figure}

Recall that the integrand \p{F3-int} includes an additional factor given  by the product
of distances between all external points. This factor breaks the $S_7$ symmetry
of the function $f^{(3)}$ down to $S_4\times S_3$. At the same time, it
simplifies the form of the graphs contributing to $F^{(3)}$  by removing the extra propagators between the external points. Some of the corresponding integrals, those which survive the light-like limit (for $x_{12}^2=x_{23}^2=x_{34}^2=x_{41}^2=0$),  already appeared in Fig.~\ref{fig:3onshell}. {Away from this limit, for arbitrary
$x_{i,i+1}^2\neq 0$}, we find only two new
conformal three-loop integrals $E$ and $H$ (for ``easy'' and ``hard'', referring to their Mellin-Barnes evaluation in Appendix~\ref{sec:three-loop-correlation function}), which are depicted in Fig.~\ref{fig:3offshell}. Thus, the complete list of integrals which we expect to find in the correlation function (away from the light-like limit)   is shown in Figs.~\ref{fig:3onshell} and \ref{fig:3offshell}.   All others graphs for $F^{(3)}$  are obtained by permuting the external points.

Here are the expressions for  the one- and two-loop ladder (box and double-box) integrals $g$ and $h$ from  Fig.~\ref{fig:g+h},    
\begin{align}
\notag
& g(1,2,3,4)  = -\frac{1}{4 \pi^2}
\int \frac{d^4x_5}{x_{15}^2 x_{25}^2 x_{35}^2 x_{45}^2}  \, , \\
\label{eq:g+h}
& h(1,2;3,4)  =  \frac{x^2_{34}}{(4 \pi^2)^2}
\int \frac{d^4x_5 \, d^4x_6}{(x_{15}^2 x_{35}^2 x_{45}^2) x_{56}^2
(x_{26}^2 x_{36}^2 x_{46}^2)}  \,,
\end{align}
and for the three-loop integrals from Figs.~\ref{fig:3onshell} and \ref{fig:3offshell}, 
\begin{eqnarray} \label{eq:14}
 T(1,2;3,4)&=&{x_{34}^2\over (-4\pi^2)^3}\int 
  \frac{ d^4x_5 d^4x_6 d^4x_7 \ x_{17}^2 }{(x_{15}^2 x_{35}^2) (x_{16}^2  x_{46}^2) (x_{37}^2 x_{27}^2  x_{47}^2)
    x_{56}^2 x_{57}^2 x_{67}^2}\ ,\nonumber \\
E(1,2;3,4) & = & \frac{x^2_{23} x^2_{24}}{(-4 \pi^2)^3}
\int \frac{d^4x_5 \, d^4x_6 \, d^4x_7 \ x^2_{16}}{(x_{15}^2 x_{25}^2 x_{35}^2)
x_{56}^2 (x_{26}^2 x_{36}^2 x^2_{46}) x^2_{67} (x_{17}^2 x_{27}^2 x_{47}^2)}
\, , \nonumber \\
L(1,2;3,4) & = & \frac{x^4_{34}}{(-4 \pi^2)^3}
\int \frac{d^4x_5 \, d^4x_6 \, d^4x_7}{(x_{15}^2 x_{35}^2 x_{45}^2) x_{56}^2
(x_{36}^2 x_{46}^2) x^2_{67} (x_{27}^2 x_{37}^2 x_{47}^2)} \, , \nonumber \\
  {(g\times h)}(1,2;3,4) &=&{x_{12}^2 x_{34}^4\over (-4\pi^2)^3}\int 
   \frac{d^4x_5d^4x_6 d^4x_7}{(x_{15}^2  x_{25}^2  x_{35}^2 x_{45}^2) 
 (x_{16}^2  x_{36}^2x_{46}^2) 
    (x_{27}^2  x_{37}^2   x_{47}^2) x_{67}^2} \ ,\nonumber\\
H(1,2;3,4) & = &   \frac{x_{41}^2 x_{23}^2
    x_{34}^2 
}{(-4 \pi^2)^3}
\int \frac{d^4x_5 \, d^4x_6 \, d^4x_7 \ x^2_{57}}{(x_{15}^2 x_{25}^2 x_{35}^2
x^2_{45}) x_{56}^2 (x_{36}^2 x^2_{46}) x^2_{67} (x_{17}^2 x_{27}^2 x^2_{37}
x_{47}^2)} \, .
\end{eqnarray}

{Summarising our analysis, we obtain the full four-point correlation 
function up to three loops as follows:
\begin{align} \label{offshc}
  G_{4}(1,2,3,4)= G^{(0)}_{4}  + \frac{2 \, (N_c^2-1)}{(4\pi^2)^{4}} \  R(1,2,3,4)   \   \left[ a  F^{(1)} + a^2 F^{(2)} + a^3 F^{(3)} + O(a^4)  \right]\,,
\end{align}
where  $F^{(1)}$ and  $F^{(2)}$ can be read off from \p{f12}, \p{integg} and \p{eq:g+h},
\begin{align}\label{}
F^{(1)} & =  g(1,2,3,4)\,, \notag
\\[2mm] \notag
 F^{(2)} & = h(1,2;3,4) + h(3,4;1,2) + h(2,3;1,4) +  h(1,4;2,3) 
 \\ & + h(1,3;2,4) + h(2,4;1,3)
  + 
 \frac12 \lr{x_{12}^2x_{34}^2+ x_{13}^2 x_{24}^2+x_{14}^2x_{23}^2}  [g(1,2,3,4)]^2\,,
\end{align}
and $F^{(3)}$ is given by the sum over all topologies }
shown in Figs.~\ref{fig:3onshell} and \ref{fig:3offshell}, symmetrised over all $S_4\times S_3$ permutations
of white and black nodes. The explicit expression for $F^{(3)}$ reads
\begin{align}\notag
F^{(3)} =&  \big[T(1,3;2,4)+ 11 \mbox{ perms}\big]
+ 
\big[E(1,3;2,4)+ 11 \mbox{ perms}\big]
\\[2mm] \notag +& \big[L(1,3;2,4)+ 5 \mbox{
  perms}\big] 
+ 
  \big[({g\times h})(1,3;2,4)+  5 \mbox{ perms}\big]
\nonumber 
\\[2mm] +&  {\textstyle \frac12}\big[  H(1,3;2,4)+ 11 \mbox{
  perms}\big],
  \label{eq:15} 
\end{align}
where ``$+11$ perms'' etc. denotes a sum over the remaining distinct
 $S_4$ permutations of external points $1,2,3,4$. The reason that the 
 hard integral $H$ comes with a factor of a half is simply that its integrand
 has the additional symmetry under $x_5\leftrightarrow x_7$ and we are summing over all distinct integrands
 with weight 1 (and then dividing by 3!). One can easily check that the
 total number of integrands on the right-hand side of \p{eq:15}  is $3!\times (12+12+6+6+12/2)=252$, as it should
 from \p{f3}. We recall that the expression for $F^{(3)}$ is defined up to
 an overall coefficient. As we show in Section~\ref{square},
 its value is fixed from the duality correlation function/amplitude to be 1. 
 

There exist identities that certain conformal off-shell integrals (but not their integrands!) satisfy for $x_{i,i+1}^2\neq 0$ (see Ref.~\cite{magic}). In
 Section~\ref{clc} we will use these identities to simplify the expression \p{eq:15} further.  It should be noted, however, that all of the above integrals diverge in the light-cone
limit $x_{i,i+1}^2\to 0$
 and need regularisation. In this limit the conformal identities do not hold anymore, so for the purpose of checking the duality correlation function/amplitude we ought to treat all integrals in \p {eq:15} as independent.  

\subsection{The square light-like limit and the four-gluon amplitude}\label{square}

Let us now verify that the proposed expression for the correlation
function \p{eq:15}  reproduces the known result for the three-loop amplitude 
in accordance  with the amplitude/correlation function duality
\p{spec}.  {In the process we will also fix the overall normalisation (the coefficient of topology 3(b)).}

To this end, we go in Eqs.~\p{offshc}-\p{eq:15}  to the light-like limit $x_{12}^2, x_{23}^2, x_{34}^2,
x_{41}^2 \rightarrow 0$ and observe that the expression on the right-hand side 
of \p{eq:15} reduces dramatically. As was mentioned in the previous subsection, all permutations of the integrals $E$ and $H$ vanish in this limit and the only non-vanishing contributions  come from 4
permutations of  $T$,  2 permutations $L$  and 2 permutations of $g\times h$.  
 Explicitly we obtain  
\begin{align} \label{eq:16}
  \lim_{x_{i,i+1}^2
  \to 0}  F^{(1)} & =  g(1,2,3,4)\,,  \nonumber\\ \notag
 \lim_{x_{i,i+1}^2 \to 0} F^{(2)} & =   h(1,3;2,4) + h(2,4;1,3) + \frac12 {x_{13}^2 
    x_{24}^2} \, [g(1,2,3,4)]^2\,, \\ \notag
  \lim_{x_{i,i+1}^2 \to 0}  F^{(3)} & =
  T(1,3;2,4)+ T(1,3;4,2)+
  T(2,4;1,3)+T(2,4;3,1)\nonumber\\&  +L(1,3;2,4) +L(2,4;1,3)+(g\times
h)(1,3;2,4) +(g\times
 h)(2,4;1,3)  \,. 
\end{align}
We recall that
the integrals in \p{eq:16} are divergent and
require regularisation. However, for the purpose of testing the duality
relation \p{spec} we will only need their integrands which are well-defined
rational functions of the coordinates of the four external points and the three 
integration points.

Let us now compare \p{eq:16} with the analogous expressions for the four-gluon
amplitude. The latter are naturally defined as functions of the on-shell particle momenta $p_i$. Going to the dual space representation, $p_i=x_i-x_{i+1}$,
the one-, two- and three-loop four-gluon amplitudes are given by \cite{magic}
\begin{align} \label{M}
M_4^{(1)} = x_{13}^2 x_{24}^2\, &  g(1,2,3,4)\,,
\notag
 \\[2mm] \notag
M_4^{(2)} = x_{13}^2 x_{24}^2 & \left[ h(1,3;2,4) + h(2,4;1,3)\right]\,, \\[2mm]  \notag 
M_4^{(3)} = x_{13}^2 x_{24}^2 &  \big[T(1,3;2,4)+ T(1,3;4,2)+
  T(2,4;1,3)+T(2,4;3,1)  
  \\[2mm] &   +L(1,3;2,4) +L(2,4;1,3) \big] \,,
\end{align}
with the integrals defined in \p{eq:g+h},  \p{eq:14}. 
Inserting \p{M} into \p{spec} and comparing   with (\ref{eq:16}) we find precise
agreement with the amplitude/correlation function prediction, as stated. 

This completes our construction of the three-loop four-point correlation function
based on the usage of the hidden $S_7$ symmetry of the integrand combined with the correlation function/amplitude duality.

\section{OPE test of the three-loop correlation function}
\label{sec:ope-test}

In this section we check that the proposed three-loop four-point
correlation function (\ref{eq:15}) agrees with the operator product expansion
(OPE). For {\it protected} scalar operators of the type  \p{O-IJ} it takes
the following general form (here we do not display the $SU(4)$ index structure):
\begin{align}\label{OPE-gen}
\cO(x_1) \cO(x_2) = \sum_{\Delta, \, S} C_{\cO \cO O_\Delta} \frac1{(x_{12}^2)^{2-\frac12(\Delta -S)}} (x_{12})_{\mu_1}\ldots (x_{12})_{\mu_S}\left[ O_\Delta^{\mu_1\ldots\mu_S}(x_2)+\dots\right] .
\end{align}
The sum on the right-hand side runs over conformal primary operators $O_\Delta^{\mu_1\ldots\mu_S}$ carrying Lorentz spin $S$ and scaling dimension $\Delta$ and over their conformal descendants shown by dots. Here we took into account 
that the scaling dimension of the operators \p{defO} equals 2 and it  is protected from quantum corrections. The contribution of each operator to the right-hand side of \p{OPE-gen}  is accompanied by the coefficient function $C_{\cO \cO O_\Delta}$, which determines the three-point correlation function 
$\la\cO(x_1) \cO(x_2) O_\Delta^{\mu_1\ldots\mu_S}(x_3)\ra$. 

In this section, we shall apply the OPE \p{OPE-gen}
to reproduce the known result for three-loop anomalous dimension of the Konishi
operator (see Eq.~\p{Konishi} below) from the four-point
correlation function (\ref{eq:15}) and to predict its three-point correlation function with the protected scalar operators \p{defO}.

\subsection{OPE expectation}
\label{sec:ope-expectation}

Let us examine the asymptotic behaviour  of the
four-point function $G_4(1,2,3,4)$ in the double short-distance limit $x_1 \rightarrow \, x_2,
\, x_3 \rightarrow x_4$. We would like to emphasise that in this section we consider the correlation function in the {\it Euclidean regime}, so that the limit $x^2_{12} \to 0$ is indeed a short-distance (coincidence) limit, $x_1 \to x_2$, and not a light-cone limit like in Section~\ref{square}.  The advantage of the double short-distance limit, 
$x_{12}^2, x_{34}^2 \to 0$, is that  we can apply the OPE  simultaneously to
 two pairs of operators in $G_4(1,2,3,4)$ and relate the leading
 asymptotics of the four-point correlation function to the contribution of a
 particular conformal primary operator, the Konishi operator.

It follows from \p{OPE-gen} that for $x_{12}^2\to 0$ the leading contribution 
to the OPE comes from the operators with the minimal twist $=(\Delta-S)$ and
the minimal Lorentz spin $S$. In particular, the identity operator $\cI$ (with $\Delta=S=0$) produces the most singular contribution $\sim 1/x_{12}^4$.
The first subleading $O(1/x_{12}^2)$ contribution originates from the gauge-invariant operators carrying zero Lorentz spin $S=0$ and tree-level (naive) scaling dimension $\Delta^{(0)}=2$.  In $\cN=4$ SYM such operators are built from the six scalars $\Phi^I$. They have the general form $C_{IJ} \tr(\Phi^I\Phi^J)$ with some coefficients $C_{IJ}$ and carry $SU(4)$ irreps from the tensor 
product  $\mathbf{6} \times \mathbf{6} = \mathbf{1} + \mathbf{15} + \mathbf{20'}$. In this way, we identify two such operators: the
 half-BPS operator $\mathcal{O}_{\mathbf{20'}}^{IJ}$, Eq.~ \p{O-IJ}, belonging
to the ${\mathbf{20'}}$ (or the symmetric traceless rank-2 representation of $SO(6)$)  and the   singlet  Konishi operator \footnote{The irrep $\mathbf{15}$ in the tensor product, i.e. the antisymmetric rank-2 tensor (the adjoint of $SU(4)$), cannot be realised as a scalar bilinear operator since $\tr(\Phi^I \Phi^J)$ is  symmetric in $I$ and $J$.}
\begin{align}\label{Konishi}
 \cK = \tr(\Phi^I\Phi^I) \,.
\end{align}
The scaling dimension of the half-BPS operator $\cO_{\mathbf{20'}}$ is protected from quantum corrections and takes its canonical value $\Delta_{\cO}  = 2$, while the Konishi operator acquires anomalous dimension:
\begin{equation}\label{K-anom}
  \Delta_\cK \, = \, 2 +
\gamma_\cK(a) = 2 + \sum_{\ell=1}^\infty a^\ell \gamma_\ell\,.
\end{equation}
The goal of our test is to extract  this anomalous dimension from the four-point correlation function up to three loops and to compare it to the known values from the literature \cite{Kon12,Rome,klovUs}:
\begin{equation}\label{expect1}
\gamma_1 \, = \, 3 \, , \qquad\quad \gamma_2 \, = \, -3 \, , \qquad\quad
\gamma_3 \, = \, \frac{21}{4}\,.
\end{equation}

Thus, restricting to the contributions of operators with naive scaling dimension up to two, the OPE of two operators $\cO(x,y)$, Eq.~\p{defO}, contains the operators $\cI$,   ${\cK}$ and  $\cO_{\mathbf{20'}}$:
\begin{align}\label{ope}
  \cO(x_1, y_1)  \; \cO(x_2, y_2)  & =     c_{\cI}  \frac{(Y_1 \cdot Y_2)^2}{x_{12}^4} \; \cI \, + c_{\cK}(a)   \frac{(Y_1 \cdot Y_2)^2}{(x_{12}^2)^{1-{\gamma_\cK}/{2}}} \;
\cK(x_2)   \nt
&+ c_\cO   \frac{(Y_1 \cdot Y_2)}{x_{12}^2} Y_{1I} Y_{2J}\, \cO_{\mathbf{20'}}^{IJ}(x_2)    \, + \ldots\,,
\end{align}
where $Y_1$ and $Y_2$ denote the $SO(6)$ harmonics depending on $y_1$
and $y_2$, respectively (see Eq.~\p{Y-def} in Appendix ~\ref{HV}).
The polynomial harmonic structure on the right-hand side of \p{ope}
is determined by the usual requirement of harmonic analyticity  and by the $SU(4)$ weights $(+4)$ at points 1 and 2 (each $Y$ carries weight $(+2)$). The dots in
\p{ope} denote the contribution of operators
of higher spin and twist. It is suppressed by powers of $x_{12}^\mu$ in the (Euclidean) limit $x_1 \rightarrow \, x_2$. The constants $c_\cI$ and $c_\cO$ 
determine the two-point and three-point functions of the half-BPS operators
$\cO_{\mathbf{20'}}$, which are known to be protected.
Therefore, the constants $c_\cI$ and $c_\cO$ do not depend on the coupling constant and are given by their tree-level expressions
\begin{align}\label{const1}
c_\cI = \frac{N_c^2 -1}{2 (4\pi^2)^2}\,,\qquad\qquad c_\cO = \frac1{2\pi^2}\,.
\end{align}
In the double short-distance limit $x_1 \rightarrow \, x_2,
\, x_3 \rightarrow x_4$ we can apply the OPE \p{ope} to the products of 
operators $\cO(x_1,y_1)\cO(x_2,y_2)$ and  $\cO(x_3,y_3)\cO(x_4,y_4)$
and express their four-point correlation function in terms of the two-point functions \footnote{To verify the second relation
in \p{o20},
it suffices to project \p{o20} with harmonics and to take the vev of both sides of \p{ope}.  } 
\begin{align} \notag
\la \cK(x_2) \, \cK(x_4) \ra & =  
\frac{d_{\cK}}{(x_{24}^2)^{2 + \gamma_\cK}}\,,
\\ 
\la \cO_{\mathbf{20'}}^{IJ}(x_2) \,
\cO_{\mathbf{20'}}^{KL}(x_4) \ra  &= \frac{ c_\cI
}{2 x_{24}^4} \left(\delta^{IK} \delta^{JL} + \delta^{IL} \delta^{JK} -\frac1{3} \delta^{IJ} \delta^{KL}\right) \label{o20} \,.
\end{align}
{Unlike the half-BPS operator $\cO_{\mathbf{20'}}^{IJ}$, the 
Konishi operator $\cK$ is not protected, so the normalisation
constants $d_\cK$ and the coefficient $c_\cK(a)$ on the right-hand side of
\p{ope} depend on the choice of the
renormalisation scheme. 
We shall adopt the convention that the Konishi operator is normalised such that $d_\cK$ keeps its free-field value:
\begin{align}\label{const2}
d_\cK = 3\,\frac{N_c^2 -1}{(4\pi^2)^2} \,, \qquad c_\cK(a) = \frac{1}{12\pi^2} + O(a)\,.
\end{align}
Below we show (see Eqs.~\p{alpha} and \p{numpre}) that 
the OPE allows us to find  $c_\cK(a)$. }

Taking into account these relations and using \p{ope} we finally obtain the
asymptotic behaviour of the four-point correlation function for $x_1\to x_2$, 
$x_3\to x_4$ as
\begin{align} \label{expect} \notag
G_4 & = \frac{y_{12}^4 y_{34}^4}{x_{12}^4 x_{34}^4} \; c_\cI^2  
\\[2mm] \notag & 
+ \frac{y_{12}^2 y_{34}^2 ( y_{13}^2 y_{24}^2 + y_{14}^2 y_{23}^2 )}{x_{12}^2 x_{34}^2 x_{24}^4} \;
\frac{c_\cO^2 c_\cI}2    \nt
&     + \frac{y_{12}^4 y_{34}^4}{x_{12}^2 x_{34}^2 x_{24}^4} \, \left( c_\cK^2(a) d_\cK  \;
u^{ {\gamma_\cK(a)}/{2}} \, - \, \frac{1}{6} \, c_\cO^2 c_\cI  \right)   \, + \, \ldots\ ,
\end{align}
where  $u= {x^2_{12}x^2_{34}}/{(x^2_{13}x^2_{24})}$ is a conformal cross-ratio
and the dots denote subleading terms. We notice using \p{const1} and  \p{const2} that the expression
in the parentheses in the last line of \p{expect} scales as $O(a)$.

Let us compare the OPE result \p{expect} with the asymptotic behaviour
of the obtained expression for the four-point correlation function
\begin{align}
G_4=G^{(0)}_4 + a\, G^{(1)}_4 + a^2 G^{(2)}_4 + a^3 G^{(3)}_4 + O(a^4)\,.
\end{align}
Replacing the tree-level expression by \p{eq:1} and the loop corrections by
\p{intriLoops} we find in the limit $x_1\to x_2$, 
$x_3\to x_4$
\begin{eqnarray}\notag
G_4 & = &  \frac{(N_c^2 -
  1)^2}{4 (4 \pi^2)^4}\  
\frac{y_{12}^4 y_{34}^4}{x_{12}^4x_{34}^4} \label{lim4p}
 \\
& + & {N_c^2 - 1\over (4 \pi^2)^4}\ \frac{y_{12}^2 y_{34}^2 ( y_{13}^2 y_{24}^2 + y_{14}^2 y_{23}^2 )}{x_{12}^2 x_{34}^2 x_{24}^4} 
 \nonumber 
 \\ & + & {2 \, (N_c^2 - 1) \over (4 \pi^2)^4}\ \frac{y_{12}^4 y_{34}^4}{x_{12}^2 x_{34}^2 } \,    
 \left(\sum_{\ell\ge 1} a^\ell F^{(\ell)} \right) \, + \, \dots \ . 
\end{eqnarray}
Arriving at this relation, we took into account that the prefactor $R$, Eq.~(\ref{eq:7}), 
scales at short distances as
$R(1,2,3,4) = y^4_{12} y^4_{34}/(x_{12}^2 x_{34}^2)+ \dots\ $.

Comparing \p{lim4p} with \p{expect} we find with the help of \p{const1} and
\p{const2} that the first two lines in the two relations coincide. Matching the
expressions in the last line we obtain the following relation
\begin{align}\label{cons}
{2 \, (N_c^2 - 1) \over (4 \pi^2)^4}\sum_{\ell\ge 1} a^\ell F^{(\ell)}(x)  =
\frac1{x_{24}^4}
\left( c_\cK^2(a) d_\cK  \;
u^{{\gamma_\cK(a)}/{2}} \, - \, \frac{1}{6} \, c_\cO^2 c_\cI  \right) +\dots \,,
\end{align}
where the dots denote terms vanishing for $x_1\to x_2$, $x_3\to x_4$. 

Expanding the right-hand side of \p{cons} in the powers of the coupling constant
and matching the coefficients  of $a^\ell$, we can obtain
the OPE prediction for the short-distance behaviour of the functions $F^{(\ell)}(x)$. 
It is convenient to write
\begin{align}\label{alpha}
c^2_\cK(a) d_\cK = \left(\frac13 + \sum_{\ell\ge 1} \alpha_\ell \, a^\ell \right)\,{N_c^2 - 1 \over (4 \pi^2)^4}\,.
\end{align}
Then, we find from \p{cons} the following relations for the functions $F^{(\ell)}(x)$
\begin{align}\notag
x_{24}^4 F^{(1)}(x) &=\frac1{12}\gamma_1 \ln u  + \frac12 \alpha_1 + \dots\,,
\\[2mm] \notag
x_{24}^4 F^{(2)}(x) &=\frac1{48} \gamma_1^2\, (\ln u)^2+\left(\frac1{12}\gamma_2+\frac14 \gamma_1\alpha_1 
\right)\ln u  + \frac12 \alpha_2 + \dots\,,
\\[2mm] \notag
x_{24}^4 F^{(3)}(x) &=\frac1{288} \gamma_1^3\, (\ln u)^3+\left(\frac1{24}\gamma_1\gamma_2+\frac1{16} \gamma_1^2 \alpha_1 \right) (\ln u)^2
\\ \label{F-as}
&\qquad  +\left(\frac1{12}\gamma_3+\frac14 \gamma_2\alpha_1 +\frac14 \gamma_1\alpha_2
\right)\ln u  + \frac12 \alpha_3 + \dots\,,
\end{align} 
where the coefficients $\gamma_\ell$ define the perturbative corrections to the anomalous dimension of the Konishi operator, Eqs.~\p{K-anom} and \p{expect1}. 
Notice that in the expansions on the right-hand sides of \p{F-as}  we have neglected the powers of the conformal cross-ratio $u$ because it vanishes
for $x_1\to x_2$ and $x_3\to x_4$.
 
The relations \p{F-as} provide powerful constraints on the form of the loop
corrections. In the next subsection we 
work out the asymptotic expansion of the obtained expression for the three-loop four-point correlation function for $x_1\to x_2$, $x_3\to x_4$ and demonstrate
that it is in agreement with \p{F-as}. Moreover, matching the coefficients  of the powers of $\ln u$, we reproduce the known result for three-loop anomalous dimension of the Konishi operator \p{expect1} and we obtain the following 
values for the coefficients $\alpha_\ell$ defined in \p{alpha} \footnote{We acknowledge the help of Volodya Smirnov in evaluating $\a_3$.} 
\begin{equation}\label{numpre}
\alpha_1 \, = \, -1 \, , \qquad \alpha_2 \, = \, \frac{3}{2}\, \zeta(3) +
\frac{7}{2} \, ,
\qquad \alpha_3 \, = \, - \left(\frac{25}{4} \, \zeta(5) + 2 \, \zeta(3) +
12  \right)\,.
\end{equation}
Notice that our reconstruction of the three-loop correlation function was limited to its leading large $N_c$ part and, therefore, the constants $\alpha_\ell$ in \p{numpre} can in principle receive corrections in $1/N_c^2$ starting from three loops.
It is worth mentioning however that the Konishi anomalous dimension $\gamma_\ell$ is known to be $N_c-$exact up to three loops \cite{Kon12,Rome,klovUs}.

{As an interesting byproduct of our study of the four-point correlation function
we can predict the three-loop result for the correlation function of two half-BPS 
operators $\cO$ and the Konishi operator
$\cK$. Indeed, it follows from the OPE, Eqs.~\p{ope} and \p{o20},  
that this correlation function looks as
\begin{align}\label{}
\langle \cO(x_1,y_1) \cO(x_2,y_2) \cK(x_3)\rangle =  \frac{C(a) \ (Y_1 \cdot Y_2)^2}{(x_{12}^2)^{1-{\gamma_\cK}/{2}}(x_{13}^2)^{1+ {\gamma_\cK}/{2}}(x_{23}^2)^{1+{\gamma_\cK}/{2}}}\,.
\end{align} 
Here the constant $C(a)=c_\cK(a) d_\cK$ takes the form 
\begin{align}\label{}
C(a) = \left(1 + 3\sum_{\ell\ge 1} \alpha_\ell \, a^\ell \right)^{1/{2}}\,{N_c^2 - 1 \over (4 \pi^2)^3}\,,
\end{align}
with  the parameters $\alpha_\ell$ given up to three loops by \p{numpre}.
We recall that the Konishi operator is defined so that it keeps its free-field normalisation, Eqs.~\p{o20} and \p{const2}.  
}

\subsection{Double short-distance limit of the correlation function}

In this subsection we work out the asymptotic expansion of the three-loop four-point correlation function $G_4$ in the limit $x_1\to x_2$, $x_3\to x_4$. According
to \p{intriLoops}, the loop corrections to $G_4$ are controlled by the functions $F^{(\ell)}$ which are given by the sum of scalar $\ell-$loop Feynman integrals
of different topology. We start by simplifying the expression for the three-loop integrals
defined in \p{eq:14}, \p{eq:g+h} and then proceed to their numerical evaluation using the Mellin-Barnes method. 

\subsubsection{Simplifying the correlation function}\label{clc}

There are various 
identities satisfied by the off-shell (that is, for $x_{i,i+1}^2\neq 0$) conformal
integrals~\cite{magic} allowing the expression~(\ref{eq:15}) to be
immediately
simplified (at the level of the off-shell integral only, not the integrand!). They follow from the fact that the conformally covariant integrals depending on four external points are functions of two conformally invariant variables, the cross-ratios
\begin{align}\label{cr}
u=\frac{x^2_{12}x^2_{34}}{x^2_{13}x^2_{24}}\,, \qquad
v=\frac{x^2_{14}x^2_{23}}{x^2_{13}x^2_{24}}\, .
\end{align}
For example, we can rewrite the one-, two- and three-loop  ladder integrals $g,h$ and $L$ from \p{eq:14}, \p{eq:g+h} as follows:
\begin{align}\label{boxes}
g(1,2,3,4) &=   \frac1{x^2_{13}x^2_{24}} \Phi^{(1)}(u,v)\,, \nt
h(1,2;3,4) & =    \frac1{x^2_{13}x^2_{24}} \Phi^{(2)}(u,v)\,, \nt
L(1,2;3,4) & = \frac1{x^2_{13}x^2_{24}} \Phi^{(3)}(u,v)\,.
\end{align}
All of these are finite four-dimensional integrals having conformal weight
one at each outer point. This weight is carried by the prefactors $1/(x^2_{13} x^2_{24})$ on the right-hand sides in \p{boxes}, which makes the functions $\Phi^{(n)}(u,v)$ conformally invariant.\footnote{The ladder functions $\Phi^{(n)}(u,v)$ are known to all orders and are expressed in terms of polylogs of maximal degree $2n$ \cite{davyPhi1}.} Similarly, the three-loop integrals $T, E$ and $H$ from \p{eq:14} can be written as 
\begin{align}\label{TEH}
T(1,2;3,4) &= \frac1{x^2_{13}x^2_{24}} T^{(3)}(u,v)\,,
\nt E(1,2;3,4) &= \frac1{x^2_{13}x^2_{24}} E^{(3)}(u,v)\,, \nt H(1,2;3,4) &= \frac1{x^2_{13}x^2_{24}} H^{(3)}(u,v)\,.
\end{align}
Notice that the prefactor $1/(x^2_{13} x^2_{24})$ and the cross-ratios 
$u,v$ are invariant under simultaneous exchange of the external points $1 \leftrightarrow 3$ and $2 \leftrightarrow 4$. Therefore, the integrals above must also be invariant under this transformation. This property is not
immediately obvious since it is not a symmetry of the integrand itself, \footnote{Except for the one-loop ladder function $g(1,2,3,4)$ which is manifestly totally symmetric under all interchanges of $x_1, \dots ,x_4$ at the level of the
integrand. }
and it
has been used in \cite{magic} to show that the two-loop ladder $h$ satisfies the ``flip identity"
\begin{equation}\label{flip}
h(1,2;3,4) \, = \, h(3,4;1,2)\ ,
\end{equation}
in addition to the manifest invariance, 
at the level of the integrand, under the exchange
$1 \leftrightarrow 2$ and $3 \leftrightarrow 4$. The three-loop ladder $L$ and the ``hard'' integral $H$, in addition to the
manifest symmetries of the integrand, satisfy the same type of
flip identity:
\begin{equation}
L(1,2;3,4) \, = \, L(3,4;1,2) \, , \qquad H(1,2;3,4) \, = \, H(3,4;1,2) \, .
\end{equation}
The ``easy'' three-loop integral $E$ has two non-obvious symmetries: the manifest
integrand $3 \leftrightarrow 4$ symmetry together with the flip
identity implies
\begin{equation}
E(1,2;3,4) \, = \, E(2,1;3,4) \, .
\end{equation}
The point labels thus come in pairs as with the other integrals so that
\begin{equation}
E(1,2;3,4) \, = \, E(3,4;1,2) \, .
\end{equation}
Finally, the property of the two-loop ladder \p{flip} has been applied in \cite{magic} to flip a two-loop subintegral in either the three-loop ladder or the tennis court, thus proving that they are identical {\it off shell} (that is, for $x_{i,i+1}^2\neq 0$)
\begin{align}\label{tcl}
T(1,2;3,4) = L(1,2;3,4)\,.
\end{align}

The identities listed above imply relations between various integrals
and allow us to simplify the three-loop expression for the correlation function
(\ref{eq:15}) leading to  
\begin{eqnarray}
F^{(3)} & = & 2 \, g(1,2,3,4) \, \left[ x^2_{12} x^2_{34} \, h(1,2;3,4) \, + \,
x^2_{13} x^2_{24} \, h(1,3;2,4) \, + \, x^2_{14} x^2_{23} \, h(1,4;2,3) \right]
\nonumber \\[2mm]
& + & 6 \left[ L(1,2;3,4) \, + \, L(1,3;2,4) \, + \, L(1,4;2,3) \right]
\nonumber \\[2mm]
& + & 4 \left[ E(1,2;3,4) \, + \, E(1,3;2,4) \, + \, E(1,4;2,3) \right]
 \nonumber \\[2mm] \label{loop3F} 
& + &  (1+1/v) H(1,2;3,4) \, + \,  (1+u/v) H(1,3;2,4) \, + \,(1+u) H(1,4;2,3)  \, . 
\end{eqnarray}
 
\subsubsection{Mellin-Barnes representation of conformal four-point integrals}

As mentioned earlier, the one-, two- and three-loop ladder integrals $g$, $h$ and $L$ (as well as the tennis court integral $T$, see \p{tcl}) are known explicitly as 
functions of the two conformal cross-ratios \cite{davyPhi1}. However, the new integrals $E$ and 
$H$ we encounter in our analysis of the three-loop correlation function have not been studied in the literature. It is beyond the scope of this paper to try to evaluate these integrals analytically. Instead, we are going to compute them numerically in the singular double short-distance limit described above, in order to exhibit the relevant logarithmic behaviour expected from \p{F-as}. In terms of the conformal ratios 
\p{cr}, the double short-distance limit $x_1\to x_2$, $x_3\to x_4$ translates into 
\begin{align}\label{limit}
u\to 0\,,\qquad v\to 1\,.
\end{align} 
We apply the standard Mellin-Barnes method \cite{MB} adapted to the conformal integrals at hand. This assumes using dual $x-$space variables instead of the more  familiar momentum space variables in the literature on Feynman integrals for amplitudes. 

We illustrate the procedure   with the simplest
example of the one-loop ladder function $g(1,2,3,4)$, Eq.~\p{eq:g+h}.
Making use of the conformal invariance in $x$ space, we may eliminate 
one of the propagators in the integral for $g(1,2,3,4)$  by
sending one point to infinity. For example, for $x_4\to\infty$ we get from 
\p{eq:g+h}
\begin{equation}\label{MB-example}
\lim_{x_4 \rightarrow \, \infty} \, x_4^2 \ g(1,2,3,4)=-
\frac{1}{4 \pi^2} \,
 \int \frac{d^4x_5}{x^2_{15} x^2_{25} x^2_{35}} \, =
\, \frac{1}{x^2_{13}} \, \Phi^{(1)}\left(\frac{x_{12}^2}{x_{13}^2},
\frac{x_{23}^2}{x_{13}^2}\right)\,.
\end{equation}
To determine the function of conformal cross-ratio $\Phi^{(1)}(u,v)$ from this
relation, it is sufficient to work out the MB representation for the integral 
in \p{MB-example} and then to replace $x_{12}^2 \to u\, x_{13}^2$ and $x_{23}^2 \to v\, x_{13}^2$. We find (the details can be found in Appendix~\ref{sec:three-loop-correlation function})
\begin{align}\label{MB-1}
\Phi^{(1)}(u,v) =-\frac{1}{4 }
\int_{-i\infty}^{i\infty} \frac{dz_1 dz_2}{(2 \pi i)^2} \, \left[\Gamma(-z_1) \Gamma(-z_2) \Gamma(1 + z_1 + z_2)\right]^2 \, 
u^{z_1}  v^{z_2}\,,
\end{align}
where the integration $z_1-$ and $z_2-$contours run along the imaginary
axis and satisfy the conditions
  ${\rm Re}\, z_1, {\rm Re} \, z_2<0$ and ${\rm Re} (1+z_1+z_2)>0$.
  
The integrations in \p{MB-1} can be performed analytically
leading to the well-known result \cite{davyPhi1} for $\Phi^{(1)}$.
For our purposes, however, we need the leading asymptotic behaviour of the MB integral \p{MB-1} in the limit \p{limit}. It can be easily obtained from \p{MB-1} by
closing the $z_1-$integration contour to the right half-plane and picking
up the coefficient of the double pole at $z_1=0$. In this way, we arrive at
\begin{align}\label{4.28}
\Phi^{(1)}(u,1) = \frac14 \ln u -\frac12 + O(u)\,.
\end{align}
The same technique can be applied to the remaining two- and three-loop
integrals  in \p{loop3F}. We have summarised some steps of our evaluation in  Appendix~\ref{sec:three-loop-correlation function}. 

Replacing the integrals in \p{loop3F} by their asymptotic expansions \p{Integrals}
we finally obtain the following result for the one-, two- and three-loop correlation functions:
\begin{eqnarray} 
 \, x_{24}^4 \,    F^{(1)}
& = &  \frac{1}{4} \, \ln u  - \frac12  \, + \, O(u) \, ,
 \nt 
x_{24}^4 \,  F^{(2)}
& = &  \frac{3}{16} \, (\ln u)^2  -   \ln u +
\frac{3}{4} \, \zeta(3) + \frac{7}{4}  \, + \, O(u) \, , \nonumber \\
\, x_{24}^4 \,   F^{(3)}
& = & \frac{3}{32} \, (\ln u)^3  - \frac{15}{16} \,
(\ln u)^2 + \left( \frac{9}{8} \, \zeta(3) + \frac{61}{16} \right)\, \ln u
 \nonumber \\ && \qquad \ \ 
- \left( \frac{25}{8} \, \zeta(5) +  \zeta(3) + 6 \right)   \, + \, O(u) \, . \label{allGs}  
\end{eqnarray}
We notice that these expressions are in perfect agreement with the OPE
prediction \p{F-as}. They allow us to reproduce the well-known result
\p{expect1}
for the Konishi anomalous dimension and to make a prediction \p{numpre} 
for the three-loop normalisation coefficients $\alpha_\ell$ defined in \p{alpha}.
 
\section{Correlation functions at higher loops}

\label{sec:four-loops}

In the previous sections, we have used the permutation symmetry of the
integrand combined with the correlation function/amplitude duality to 
construct the three-loop four-point correlation function. It is fairly straightforward 
to extend our analysis to higher loops. As an example, we consider in
this section the  four-loop four-point correlation function.

From (\ref{eq:8}) we can write the four-loop integrand for the
four-point correlation function  
\begin{align}\label{G8}
      G_{8;4}^{(0)}(1,\dots, 8) &= \frac{2 \, (N^2_c-1)}{(4 \,
\pi^2)^{8}} \times  \cI_{8} \times
 f^{(4)}(x_1, \dots, x_{8})\,,
\end{align}
where $f^{(4)}(x_1, \dots, x_8)$ is conformally covariant with
weight $(+4)$ at each point and is $S_8-$ invariant. According to
\p{eq:10}, this function has the following form 
\begin{align}\label{f8}
f^{(4)}(x_1 \dots, x_{8})= { P^{(4)}(x_1, \dots
   x_{8}) \over \prod_{1\leq i<j \leq 8}   x_{ij}^2}\ ,
\end{align}
where $P^{(4)}$ is a homogeneous polynomial in $x_{ij}^2$ of degree $12$,  having conformal  weight $(-3)$ at each point.

Let us analyse the possible choices for the polynomial $P^{(4)}(x_1, \dots,
x_8)$. As at three loops,  it is convenient to use a diagrammatic representation of $P^{(4)}$ as a graph with vertices corresponding to the
points $x_i$  and the dashed lines denoting the terms $x_{ij}^2$.
The fact that $P^{(4)}(x_1, \dots, x_8)$ has weight $(-3)$ at each point implies that there are exactly
three lines attached to every vertex. In graph theory terminology, such graphs
are known as regular multi-graphs of degree 3.~\footnote{In graph theory
  a ``graph'' should have only 1 line between any two points, for
  multi-graphs we are allowed multiple lines between two points. Here we
  have such topologies, they simply correspond to allowing $x_{ij}^4$
  in the numerator.}  Thus the problem
of analysing all possibilities for the polynomial $P^{(4)}(x_1, \dots, x_8)$ is equivalent to the mathematical problem of counting all non-isomorphic (connected and disconnected) multi-graphs with $8$ vertices of degree 3. 
By doing this, with the help of~\cite{multigraph}, we find the total number of independent possibilities for $P^{(4)}$ is $32$.

Thus, using the general properties of \p{G8} alone, we find that the integrand of the 
four-loop correlation function is parametrised  by 32 unknown coefficients (one per each inequivalent contributing topology). To determine these coefficients
we have to impose additional conditions on the function $f^{(4)}(x_1, \dots, x_{8})$. 
So far our considerations were valid in $\cN=4$ SYM with gauge group $SU(N_c)$ for any number of colours $N_c$. To formulate the above conditions, we plan to use  the correlation function/amplitude duality, which is valid in the planar limit only. Therefore we shall restrict ourselves to the planar limit of the correlation function. As at three loops, the coefficients in front of the 32 topologies
can be determined from the correlation function/amplitude duality, together with the known four-loop four-gluon amplitude result~\cite{4loopMHV}. 
  

Comparison with the four-particle planar amplitude uniquely determines the four-loop integrand of 
the four-point correlation function to be
\begin{align}\label{G8-sum}
&  G_{8;4}^{(0)}(1,\ldots,8) = {2
    (N_c^2-1) \over (4 \pi^2)^8}  \times \cI_8 \times
\ {\sum_{\sigma \in S_8}  \left[ P_A(x_{\sigma(i)})+P_B(x_{\sigma(i)})-P_C(x_{\sigma(i)})\right] \over \prod_{1\leq i<j \leq 8}
    x_{ij}^2}\ ,
\end{align}
where the contributions from topologies $A,B$ and $C$ are given by
\begin{align}\label{P-abc}
P_A(x_1, \dots, x_8)&={1\over 24} x_{12}^2 x_{13}^2 x_{16}^2 x_{23}^2 x_{25}^2 x_{34}^2 x_{45}^2 x_{46}^2 x_{56}^2 x_{78}^6\,,  \nonumber\\
P_B(x_1 \dots, x_8)&={1\over8}x_{12}^2 x_{13}^2 x_{16}^2
   x_{24}^2 x_{27}^2 x_{34}^2 x_{38}^2 x_{45}^2 x_{56}^4 x_{78}^4 \,, \nonumber\\
P_C(x_1 \dots, x_8)&={1\over 16} x_{12}^2 x_{15}^2 x_{18}^2 x_{23}^2 x_{26}^2 x_{34}^2
   x_{37}^2 x_{45}^2 x_{48}^2 x_{56}^2 x_{67}^2 x_{78}^2 
\ .
\end{align}
Here the multiplicative fraction on the right-hand side are inserted to 
ensure that each term is only counted once in the sum over permutations in
\p{G8-sum}. The fact that $P_A$, $P_B$ and
$P_C$ come with equal weight and $P_C$ with a minus sign follows from
comparing with the known result for the four-loop four-gluon  amplitude \cite{4loopMHV} where all integrals appear with equal or opposite weight.

\begin{figure}[h!]
\psfrag{i1}[cc][cc]{$\sigma_1$}
\psfrag{i2}[cc][cc]{$\sigma_2$}
\psfrag{i3}[cc][cc]{$\sigma_3$}
\psfrag{i4}[cc][cc]{$\sigma_4$}
\psfrag{i5}[cc][cc]{$\sigma_5$}
\psfrag{i6}[cc][cc]{$\sigma_6$}
\psfrag{i7}[cc][cc]{$\sigma_7$}
\psfrag{i8}[cc][cc]{$\sigma_8$}
\psfrag{a}[cc][cc]{$P_A$}
\psfrag{b}[cc][cc]{$P_B$}
\psfrag{c}[cc][cc]{$P_C$}
\vspace*{5mm}
\centerline{\includegraphics[height=48mm]{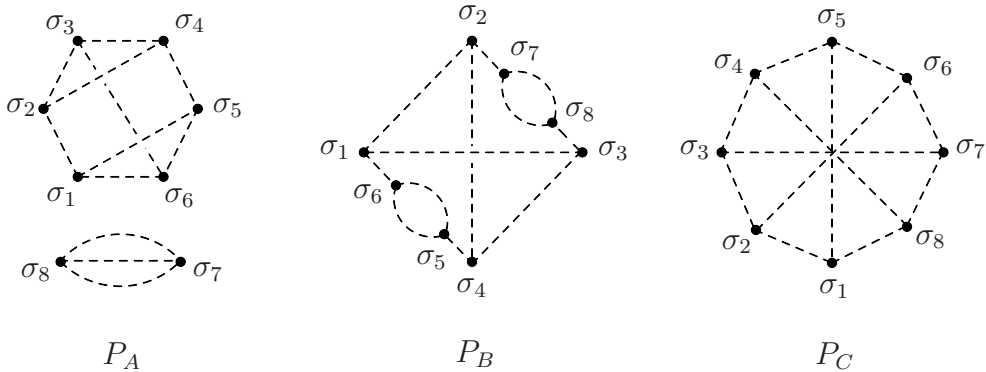}}
\caption{\small  Diagrammatic representation of the three 
conformally covariant polynomials defined in \p{P-abc}. 
These are the only $S_8$ topologies occurring in the four-loop four-point
correlation function in the planar limit.
} 
\label{fig:s8topamp} 
\end{figure} 

\begin{figure}[h!]
\psfrag{i1}[cc][cc]{$\sigma_1$}
\psfrag{i2}[cc][cc]{$\sigma_2$}
\psfrag{i3}[cc][cc]{$\sigma_3$}
\psfrag{i4}[cc][cc]{$\sigma_4$}
\psfrag{i5}[cc][cc]{$\sigma_5$}
\psfrag{i6}[cc][cc]{$\sigma_6$}
\psfrag{i7}[cc][cc]{$\sigma_7$}
\psfrag{i8}[cc][cc]{$\sigma_8$}
\psfrag{a}[cc][cc]{$f_A$}
\psfrag{b}[cc][cc]{$f_B$}
\psfrag{c}[cc][cc]{$f_C$}
\vspace*{5mm}
\centerline{\includegraphics[height=53mm]{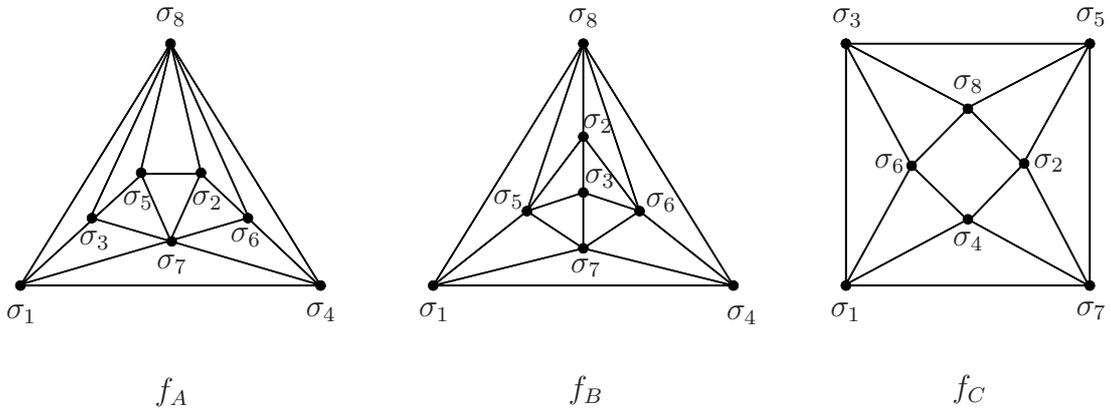}}
\caption{\small  Diagrammatic representation for the function $f^{(4)}(x_i)$, Eq.\p{f8}, corresponding to the polynomials depicted in Fig.~\ref{fig:s8topamp}. 
{To simplify the figures we did not show the dashed lines corresponding to
$x_{\sigma_7\sigma_8}^4$ for $f_A$ and to $x_{\sigma_5\sigma_6}^2 x_{\sigma_7\sigma_8}^2$ for $f_B$.}
To obtain the four-loop integrand of $F^{(4)}(x_i)$ from these graphs, one should  
separate the eight vertices into four external and four internal points and remove the 
propagators connecting the external points only. } 
\label{fig:8f} 
\end{figure} 

The three $S_8$ topologies of the polynomial $P^{(4)}$ 
which appear on the right-hand side of \p{G8-sum} are shown graphically  in Figure~\ref{fig:s8topamp}. The corresponding contributions to the function $f^{(4)}$, Eq.~\p{f8}, are shown
in Figure~\ref{fig:8f}.

As at three
loops, all the standard integrand graphs known from the four-point
amplitude are obtained from these by choosing 4 of the 8 vertices to
be external vertices, and deleting all edges between these vertices
(corresponding to external propagators). 

{We would like to emphasise that the new permutation symmetry of the integrand function $f^{(4)}$ explains in part why the various four-loop integral topologies appearing in the amplitude come with coefficients $(\pm1)$. The total of 8 such integrals from Ref.~ \cite{4loopMHV} fall into the three classes $A,B,C$ described above. Then, the symmetry of $f^{(4)}$ implies that all the integrals from the same class must come with equal coefficients. However, this symmetry does not explain the relative coefficients between the three classes (for more comments see Sect.~\ref{Conc}). }

{The only additional thing we need to check is that none of the 32
  topologies  completely vanish in the light-like
  limit (in which case we could clearly not determine them by this
  method).
 To prove this it is enough to show  that for any
allowed numerator topology we can choose vertices in such a way that $x_2$ is
not adjacent to $x_1$ or  $x_3$ and $x_4$ is not adjacent to $x_3$ or
$x_1$. To do this simply choose two adjacent points (labeled $x_1$
and $x_3$) and consider the set of all points adjacent to either $x_1$
or $x_3$. There can be at most four such points (since the (multi-)graph has
degree 3), so we can choose these
to be $x_5,x_6,x_7,x_8$. Then the remaining points $x_2$ and $x_4$ cannot be adjacent to $x_1$ or $x_3$. We have thus shown that for any
topology there exists at least one permutation which will be non-vanishing in the
light-like limit. Thus the four-point amplitude  limit is indeed
enough to completely determine the
correlation function. \footnote{
There is in fact a nice graph theoretical description of whether
contributions to $f^{(\ell)}$ completely vanish in the light-like limit or not. In order for a contribution from $f^{(\ell)}$ to be
non-vanishing in the  light-like limit, it needs to
contain the term $1/(x_{12}^2x_{23}^2 x_{34}^2x_{14}^2)$,  cancelling
corresponding terms in the prefactor of~(\ref{integg}). If these are
not present, the prefactor will force this contribution to vanish. Now
any $S_{4+\ell}$ topology will contain such a term {\em as long
  as the corresponding graph (describing $f^{(\ell)}$ and not the
  numerator polynomial $P^{(\ell)}$) contains a 4-cycle} (i.e. four points with
  vertices connected in a square configuration). 
This is the case for $\ell=4$ and $\ell=5$. }
}

{The fact that the four-loop correlation function is compatible at all
with the correlation function/amplitude duality is itself a highly
non-trivial consistency requirement. It would be interesting to test
this result against the OPE prediction as was done at three loops
in Section~\ref{sec:ope-test}.}

We can easily continue and do the same analysis at five loops based on the
known five-loop four-gluon amplitude integrands~\cite{5loops}.

\section{Conclusions}\label{Conc}

In this paper we have obtained a general  expression for the integrand of the
three-loop, four-point 
correlation function of stress-tensor multiplets for  any gauge group,
with four undetermined coefficients. In the planar $SU(N_c)$ theory we
determine these coefficients straightforwardly and uniquely by
comparing with the 
corresponding  four-gluon scattering amplitude and using the amplitude/correlation function
duality. To test this result we have compared with the expectations of
the OPE in a certain short-distance limit and as a byproduct obtained the three-point
function $\vev{\cO \cO \cK}$ at three-loops.  We have also obtained the four-point 
correlation function  at
four loops, albeit without the corresponding OPE test.  One of the
keys to the derivation was the discovery of an unexpected symmetry mixing internal and external points in the
integrand.

There are a
number of further comments we wish to make. 
Firstly, in this paper we have been mostly interested in deriving
correlation functions. In particular, we have emphasised the presence of this additional symmetry
in the {\em correlation function}, where it manifests itself as a simple permutation symmetry. However, the techniques and observations
in this paper have a great deal to say about higher-loop {\em amplitudes}
as well. Firstly, note that a part of the $S_{4+\ell}$ symmetry
remains intact after having taken the light-like limit, and thus gives
a new hidden symmetry for the four-point amplitude integrand via the
amplitude/correlation function duality. Indeed, this
hidden, broken $S_{4+\ell}$ symmetry
 explains why many of the amplitude graphs appearing in higher-loop
 four-point amplitudes have equal coefficients. It does not,
 however,  fix the overall
 coefficients. Still, we can get more from the amplitude/correlation
 function duality by observing that the correlation function at $\ell$ loops contains
 contributions from products of lower-loop amplitudes as well as the
 $\ell-$loop amplitude itself (see for example~(\ref{spec})). Now, if we rewrite the duality relation \p{du} as $\lim (G_4/G^{(0)}_4)^{1/2} = A_4/A^{(0)}_4$, we realise that such product terms of non-planar topology should not appear on the right-hand side. This gives a consistency condition which fixes the coefficients of certain
topologies in the correlation function, which in turn then feed back
in and completely
determine many of the coefficients of the $\ell-$loop amplitude itself. For
example, at three loops, the coefficient in front of topology of Fig.~3(b) and
hence the entire correlation function is
{\em completely fixed} by~(\ref{spec}) together with the one- and two-loop
amplitudes and the planarity selection rule we discuss below which
eliminates the other topologies. Thus the three-loop amplitude can be
completely derived from lower-loop amplitudes via the
amplitude/correlator duality.  At four loops, the coefficients of $f_A$ and $f_B$ in
Fig.~\ref{fig:8f} are completely determined by the contributions of
products of lower-loop amplitudes, these in turn then determine much
(but not all) of the four-loop amplitude. Specifically, the coefficient
in front of $f_C$ is not fixed by this method. On the amplitude side $f_C$ contributes
to the two amplitude integrals $f_2$ and $d_2$ in~\cite{4loopMHV}. Thus, we
find we can derive all terms in  the four-loop amplitude from lower-loop amplitudes, {\em except} the (equal) coefficients of $f_2$ and $d_2$.   Interestingly, the coefficients of
all  four-loop amplitude
integrals equal $(+1)$ (these are the coefficients which are fixed by lower loops by
 this method) {\em except} for the 
$f_2$ and
 $d_2$ coefficients which are then equal to $(-1)$. We currently have no
 explanation for this (see \cite{Cachazo:2008dx} for an alternative explanation).

The second comment concerns planarity. We can make an intriguing observation concerning
the topology of the graphs contributing to $f^{(\ell)}$ in the large $N_c$
limit. These are the graphs
given at two loops in in Fig.~\ref{1+2-loop}(right panel), at three loops in
Fig.~\ref{fig2}(b) and at four loops in Fig.~\ref{fig:8f}. All these
graphs (with the notable exception of one loop) are planar, i.e. they can all
be drawn on a sheet of paper without any lines crossing, and
furthermore all other possible contributions to $f$ are from non-planar graphs. If this
observation 
continues to higher loops, it would provide
a potentially simple and powerful way of greatly reducing the number of graphs one need
consider in the large $N_c$ limit. However it is not immediately clear why these  
graphs for $f^{(\ell)}$ should be planar (recall that they do not correspond directly
to the correlator or the amplitude, but require the attaching of various
external propagators). 

What we {\em can} argue however is that the graphs contributing
to the Born-level $(4+\ell)-$point correlation function $G_{4+\ell;\ell}^{(0)}$ (which
defines for us the 
$\ell$-loop integrand) should be planar in the standard
sense. This planar correlation function is related to $f^{(\ell)}$ via~(\ref{eq:8}). Picking a particular
component correlation function (for example, choosing the $SU(4)$ channel $y_{13}^4
y_{24}^4$) and using~(\ref{eq:6}, \ref{eq:7})),   we find that 
\begin{equation}
  \label{eq:2}
  f^{(\ell)} \times x_{12}^2 x_{23}^2 x_{34}^2 x_{14}^2
\end{equation}
should produce only  planar diagrams. Indeed, one finds that insisting
on this property at three loops uniquely selects topology 3(b) from the
four topologies in Fig.~\ref{fig2}. The coefficient of topology
3(b) will not be determined by this method, but the OPE results in
Section~\ref{sec:ope-test} are enough to fix this remaining
coefficient.

We may thus view the above arguments as an alternative derivation of the
three-loop correlation function to the one presented in the main text, with the
additional benefit that this derivation requires {\em no} input from the 
amplitude/correlation function conjecture at all.
This method however appears to be less useful at higher loops where one is
left with more and more unfixed coefficients. 
 
{ Another comment we would like to make concerns the actual functional dependence of the correlation function. What we have done so far is to present the set of Feynman integrals which, once computed, will determine the single function of two conformal cross-ratios entering in $F^{(4)}$ in \p{integg}. The explicit functions for the ladder diagrams in \p{boxes} (and for the tennis court $T$ in \p{TEH}) are known \cite{davyPhi1} and they are expressed in terms of standard polylogs. It would be very interesting to try to evaluate the two new three-loop integrals $E$ and $H$, Eq.~\p{TEH}, that we predict. An intriguing question is if some new types of special functions will appear there and whether the so-called ``maximal transcendentality" rule will be respected. We may compare this problem to that of the evaluation of the so-called six-point ``remainder function" \cite{dual21,dual22,Anastasiou:2009kna,DelDuca:2010zg}. Recently the result of the Feynman integral two-loop calculations  has been  put in a remarkably simple form \cite{spradGonch} involving only classical polylogs. We recall that the six-point remainder function depends on three variables, while the four-point correlation function depends on two variables. Therefore, it seems that with the correlation function we have a better chance to probe this interesting mathematical problem at higher loops. }

\section*{Acknowledgements}

{We are grateful to Claude Duhr, Johannes Henn and  Valya Khoze and especially to  Volodya Smirnov  for a
number of enlightening discussions. B.E. and P.H. acknowledge support
by STFC under the rolling grant ST/G000433/1. G.K. and E.S. acknowledge the warm hospitality of the KITP at Santa Barbara, where part of this work was done. }

\appendix

\section{Conventions for the harmonic variables}\label{HV}

Throughout the paper we use two sets of harmonic variables, one for projecting $SU(4)$ indices, $g^b_A =( \delta^b_a,y^b_{a'})$ (with the $SU(4)$ indices split as $A=(a,a')$) and the other for $SO(6)$ indices, $Y^I$. They are related to each other as follows 
\begin{align}\label{Y-def}
Y_I = \frac1{\sqrt{2}} (\Sigma_I)^{AB}  \epsilon_{ab}\, g^a_A g^b_B\,.
\end{align}
 Here $\Sigma_I$ are building blocks in the expression for the Dirac matrices in six-dimensional Euclidean space, see Eq.~(B.17) in Ref.~\cite{Belitsky:2003sh}. They  satisfy
\begin{align}
  \sum_{I=1}^6 (\Sigma_I)^{AB} (\Sigma_I)^{CD} =  \frac12 \epsilon^{ABCD}\,.
\end{align}
Using the definition \p{Y-def} we find  
\begin{align}
  (Y_1\cdot Y_2) \equiv   \sum_I (Y_1)_I (Y_2)_I = \frac12\epsilon_{ab} \epsilon^{a'b'}
(y_{12})_{a'}^a (y_{12})_{b'}^b \equiv  y_{12}^2\,,
\end{align}
where $(y_{12}) _{a'}^a =(y_{1} -y_{2})_{a'}^a$.

The same matrices $(\Sigma_I)^{AB}$ relate the two sets of scalar fields:  the six pseudo-real   fields $\phi^{AB}=-\phi^{BA}$  (with $A,B=1,\ldots,4$) and the six real fields $\Phi^I$ (with $I=1,\ldots,6$)
\begin{align}
\phi^{AB} = \frac1{\sqrt{2}} (\Sigma_I)^{AB} \Phi^I\,,\qquad
\bar\phi_{AB} = \frac12\epsilon_{ABCD} \phi^{CD} 
\end{align}
or equivalently
\begin{align}
Y_I \,\Phi^I = \epsilon_{ab}\, g^a_A g^b_B\, \phi^{AB}\,.
\end{align} 
In this paper we consider the correlation functions of half-BPS scalar operators
\begin{align}\notag
\mathcal{O}^{IJ} &= \tr\left(\Phi^I \Phi^J\right) - \frac16 \delta^{IJ} \tr\left(\Phi^K \Phi^K\right) \,,
\\
\mathcal{O}^{AB,CD}  &
= \tr\left(\phi^{AB} \phi^{CD}\right) - \frac1{6} \epsilon^{ABCD} \tr\left(\phi^{EF} \bar \phi_{EF}\right) \,.
\end{align}
They are related to each other as
\begin{align}
\mathcal{O}^{AB,CD} = \frac12 (\Sigma_I)^{AB}(\Sigma_J)^{CD} \mathcal{O}^{IJ}\,.
\end{align}

\section{Maximal chiral half-BPS nilpotent invariants}
\label{sec:an-s_n-invariant}    

In this appendix we construct the nilpotent superconformal invariants $\cI_n$ appearing in the expression for the correlation function \p{eq:8} and explain their properties announced in Section~\ref{sect.2.2}. Similar invariants have been extensively used in the past, also in the framework of the Lagrangian insertion procedure, but for four-point correlation functions of  $\cN=2$ hypermultiplets \cite{ESS,Arutyunov:2003ad}. There the  outer points of the correlation function are Grassmann analytic (half-BPS) while the insertion points are chiral (as the $\cN=2$ SYM Lagrangian itself).  The case $\cN=4$ is rather different because both the outer and insertion points are Grassmann analytic.   

A preliminary study of such $\cN=4$ nilpotent invariants was undertaken in \cite{EHW}.  Later on a systematic method for building the $PSU(2,2|4)$ invariants was developed,   the supergroup formalism of~\cite{Heslop:2001dr,Heslop:2002hp,Heslop:2003xu}. This technique is universal, but it is somewhat
technically  involved. The type of invariant we need here is very special -- it has the maximal allowed Grassmann degree for a given number of points (see Eq.~\p{eq:6}) and we are able to present a simple alternative construction  which makes all properties of the invariant manifest.

\subsection{General properties of the $\cN=4$  half-BPS superconformal invariants}

The correlation functions $\vev{\cT(1) \ldots \cT(n)}$  are defined in analytic superspace, which involves half of the Grassmann variables, four chiral $\rho^a_\a$ (see \p{anth}) and four anti-chiral $\bar\rho^\da_{a'}$. Part of the superconformal group  $SL(4|4)$ (the  complexification of $PSU(2,2|4)$), the chiral Poincar\'e supersymmetry $Q$ and the anti-chiral special conformal supersymmetry $\bar S$ act on the chiral odd variables $\rho$ as shifts:
\begin{align}\notag
\delta_Q \rho{}_\alpha^a &= \epsilon_\alpha^a+\epsilon_\alpha^{a'} y{}_{a'}^a \,,
\\
\delta_{\bar S} \rho{}_\alpha^a &= x{}_\alpha^{\dot\alpha}\lr{\bar\xi_{\dot\alpha}^a+\bar\xi_{\dot\alpha}^{a'} y{}_{a'}^a }  \,,  \label{q+s}
\end{align} 
and analogously for the variables $\bar\rho$ and generators $\bar Q$ and $S$. The total number of odd parameters $\ep,\bar\xi$  in \p{q+s} is 16. Using this freedom, we can gauge away four sets of analytic odd variables, for example,
\begin{align} 
&\mbox{$Q+\bar S$ gauge:} \qquad \rho_1=\rho_2=\rho_3=\rho_4=0 \label{qbs} \,, 
\end{align} 
as well as\footnote{ In the complexified description $\rho$ and $\bar\rho$ are treated as independent, so the two gauges can be fixed separately.} 
\begin{align} 
&\mbox{$\bar Q+S$ gauge:} \qquad \bar\rho_1=\bar\rho_2=\bar\rho_3=\bar\rho_4=0 \label{bqs}  \,.
\end{align}
If we restrict ourselves to the four-point correlation function $\vev{\cT(1) \ldots \cT(4)}$, then the entire dependence on the odd variables is completely fixed by superconformal symmetry: given the $\rho=\bar\rho=0$ component \p{cor4loop},  the global supersymmetry transformation which leads to the gauges  \p{qbs}, \p{bqs} unambiguously restores the dependence on $\rho,\bar\rho$.

However, with $n \geq 5$ points we can have {\it nilpotent} superconformal invariants. Indeed, the gauges  \p{qbs}, \p{bqs} completely exhaust the power of the odd part of $SL(4|4)$; any remaining $\rho_i$ (or $\bar\rho_i$) with $i=5,\ldots,n$ will automatically be supersymmetric  invariant. Further,  the $\mathbb{Z}_4$  centre of  the R symmetry group $SU(4)$ acts on the odd variables by rescaling: $\rho \to \omega \rho$ and $\bar\rho \to \omega^{-1} \bar\rho$ with $\omega^4=1$. Thus, we have the following types of $\mathbb{Z}_4$ invariants: chiral $(\rho)^{4k}$,  anti-chiral $(\bar\rho)^{4k}$  and mixed $(\rho\bar\rho)^{k}$,  or combinations thereof. In each case the  Grassmann degree cannot exceed the value $4n-16$ corresponding to the number of odd variables left at our disposal after fixing the gauges \p{qbs} and \p{bqs}. We shall call {\it maximal} the nilpotent invariants  which involve all the available odd variables. For instance,  the maximal chiral invariant is of the form $(\rho)^{4(n-4)}$.  Notice that such invariants are unique, up to an overall factor depending only on $x$ and $y$.  The dependence on the harmonic variables $y$ is further restricted by harmonic analyticity (i.e., irreducibility under $SU(4)$), which allows only polynomial dependence of a fixed degree, in our case of degree four (matching the $U(1)$ charge of the stress-tensor multiplet $\cT$).

Now, the subject of this paper are the four-point correlation functions $\vev{\cT(1) \ldots \cT(4)}$. As explained above, their dependence on the odd variables is obtained by transforming the bosonic variables $x,y$ under conformal supersymmetry with the parameters which have been used to achieve the gauges  \p{qbs}, \p{bqs}. Such transformations always create pairs $\rho_i \bar\rho_i$. Therefore, if we decide to eliminate the anti-chiral variables by fixing the gauge \p{bqs}, we automatically eliminate the chiral ones as well. 

The situation is different for $n \geq 5$: There, even in the gauges  \p{qbs}, \p{bqs}, we can have purely chiral (and anti-chiral) or mixed nilpotent invariants. But again in the context of this paper, for the purpose of Lagrangian insertions into the four-point correlation function (see \p{eq:3}), the relevant part of the higher-point correlation functions are the components in which the  $\cN=4$ SYM {\it chiral} Lagrangian appears at $n-4$ points.  This Lagrangian belongs to the purely chiral sector of the stress-tensor multiplet, see \p{Tch}. For this reason we may ignore the $\bar\rho$ dependence of the correlation function, both for $n=4$ (when it is fixed by supersymmetry) and $n>4$ (when it is irrelevant for our purposes). This justifies our decision to set all $\bar\rho_i=0$, with $i=1,\ldots,n$. 

So, according to the insertion formula \p{eq:3}, we are interested in the $n-$point correlation function \p{nilG},  $G_{4+\ell; \ell}^{(0)} \sim (\rho_5)^4 \ldots (\rho_{4+\ell})^4 + \ldots $, where the dots denote the superconformal completion to a full $n-$point purely chiral nilpotent invariant of the {\it maximal} type. Consequently, its dependence on the odd variables is uniquely determined by supersymmetry. In the next subsection we construct such an invariant and show that the only freedom in it is a factor depending only on $x$. Most importantly, our construction makes the permutation symmetry $S_n$ with respect to all points manifest. 

\subsection{Constructing the maximal nilpotent invariant}

Consider the following  chiral Grassmann integral
\begin{align}\label{0.1}
\cI_n(x_i,y_i,\rho_i) = \int d^4 \epsilon \,  d^4 \epsilon'  d^4\bar\xi  \, d^4\bar\xi' \  \prod_{i=1}^n\delta^{(4)}
\lr{\rho_i{}_\alpha^a - \lr{\epsilon_\alpha^a+\epsilon_\alpha^{a'} y_i{}_{a'}^a}
- x_i{}_\alpha^{\dot\alpha}\lr{\bar\xi_{\dot\alpha}^a+\bar\xi_{\dot\alpha}^{a'} y_i{}_{a'}^a }}\,.
\end{align} 
It depends on $n$ points $(x_i,y_i,\rho_i)$ in the chiral analytic
superspace and has the maximal Grassmann degree $4n-16$.
The invariance of the integral under shifts of the integration variables clearly implies invariance
of $\cI_n$ under $Q$ and $\bar S$ supersymmetry \p{q+s}.   

Let us examine the properties of $\cI_n$ under the conformal inversion 
$x_i \to x_i^{-1}$, $\rho_i \to  x_i^{-1}\rho_i$:
\begin{align}\notag
\cI_n(x_i^{-1},y_i,x_i^{-1}\rho_i) 
&=  \int d^4 \epsilon\, d^4 \epsilon'   d^4\bar\xi  \, d^4\bar\xi' \ \prod_i  (x^2_i)^{-2} \delta^{(4)} 
\lr{\rho_i  - x_i\lr{\epsilon +\epsilon' y_i}- \lr{\bar\xi +\bar\xi' y_i }}\,.
\end{align}  
Exchanging the integration variables $\epsilon \leftrightarrow \xi$ and $\epsilon' \leftrightarrow \xi'$ we arrive at 
\begin{align}
\cI_n(x_i^{-1},y_i,x_i^{-1}\rho_i) =  {\prod_{k=1}^n \lr{x_k^2}^{-2}}\times \cI_n(x_i,y_i,\rho_i) \,.
\end{align}
In addition, it is easy to show that $\cI_n$ is also invariant under
shifts $x_i \to x_i + \delta$.
Since special conformal transformations are a combination of translations and inversions, $K=I P I$, we see that $\cI_n(x_i,y_i,\rho_i)$ is conformally covariant with weights $(-2)$ at each point. 

The R symmetry group $SU(4)$ can be treated in complete analogy with the conformal group $SU(2,2)$ (both being real forms of $SL(4,\mathbb{C})$). 
Its action can be reduced to a combination  of  translations of $y{}^a_{a'}$
and inversion in the group space,
\begin{align}\label{}
y^a_{a'} \to (y^{-1})^{a'}_a\,, \qquad \quad
\rho^a \to \rho^a  (y^{-1})^{a'}_a \,.
\end{align} 
We verify that $\cI_n$ is covariant under $SU(4)$ as well,
\begin{align}\label{}
\cI_n(x_i,y_i^{-1} ,\rho_i \, y_i^{-1}) =  {\prod_{k=1}^n \lr{y_k^2}^{-2}}\times \cI_n(x_i,y_i,\rho_i) \,.  
\end{align}

We remark that, being  a chiral object,  $\cI_n(x_i,y_i,\rho_i)$   is not invariant under the anti-chiral part of Poincar\'e supersymmetry which acts only on the $x$'s, $\delta_{\bar Q} x^{\da\a} = \rho^\a\bar\ep^\da$. The same applies to the chiral half of special conformal supersymmetry $S=I\bar Q I$. 
In fact, in the gauge \p{bqs} all the parameters $\bar\ep$ and $\xi$ are fixed and $\bar Q+S$ supersymmetry is completely broken. If needed, the dependence on $\bar\rho$ and with it the  $\bar Q+S$ invariance may be restored by undoing the gauge fixing, but this is not necessary here. 

To summarise, the nilpotent object $\cI_n$ constructed in \p{0.1} is:\begin{itemize}
\item chiral and Grassmann analytic of maximal Grassmann degree $4(n-4)$;
\item  invariant under $Q+\bar S$ supersymmetry;
\item  covariant under $SU(2,2)\times SU(4)$ with conformal weight $(-2)$ and $SU(4)$ weight $(+4)$ at each point;
\item  harmonic analytic, i.e. polynomial in the $y$'s of degree four in each $y_i$ matching the $SU(4)$ weight;
\item fully symmetric under the exchange of any pair of superspace points $(x_i,y_i,\rho_i)$. 
\end{itemize}

What is the freedom left in this invariant? As explained earlier, the $\rho-$dependence of $\cI_n$ is already fixed by supersymmetry. We could modify $\cI_n$ by multiplying it by a bosonic function $f(x_i,y_i)$. For our purposes we need an invariant with $SU(4)$ weights $(+4)$ at each point, to match the properties of the stress-tensor multiplet $\cT$. The invariant $\cI_n$ already has the required weight, therefore the function $f(x_i,y_i)$ must depend on harmonic cross-ratios (the $y-$analogs of the space-time cross-ratios  \p{cr}). However, the latter inevitably introduce harmonic singularities (poles in $y_{ij}^2$) which are forbidden by harmonic analyticity. We conclude that the only freedom left in the invariant $\cI_n$ is a function $f(x_i)$. It must carry conformal weight $(+4)$ at each point in order for $\cI_n(x_i,\rho_i,y_i) \times f(x_i)$ to match the weight $(+2)$ of the  $\cT$'s. Finally, the complete crossing symmetry of the correlation function  $\vev{\cT(1) \ldots \cT(n)}$, combined with the point permutation symmetry of $\cI_n$, imply that $f(x_i)$ must have full symmetry $S_n$. This proves our claim made in \p{eq:8}.

\subsection{Gauge-fixed form}

Evaluating the Grassmann integrals in \p{0.1} in general is cumbersome and leads to a complicated Jacobian involving $x$ and $y$. The calculation is greatly facilitated by choosing the $(Q+\bar S)$ gauge \p{qbs}, $\rho_1=\ldots=\rho_4=0$. In addition, we may use part of the conformal and $SU(4)$ symmetry (translations and inversion) to fix the  
\begin{align}\label{Kga}
\mbox{$SU(2,2)\times SU(4)$ gauge:} \qquad    
x_2, y_2\to\infty \,, \quad x_3=y_3=0\,. 
\end{align}
The rest of $SU(2,2)\times SU(4)$ allows us to bring the remaining $x$ and $y$ variables to the diagonal $2\times2$ matrix form \cite{Heslop:2002hp,Dolan:2004mu}
\begin{align}\label{0.16}
x_{14} x_1^{-1} = \left[\begin{array}{cc}z & 0  \\0 & \bar z \end{array}\right]\,,\qquad
 y_{14} y_1^{-1} = \left[\begin{array}{cc}w & 0  \\0 & \bar w \end{array}\right]\,.
\end{align}
In the gauge \p{Kga}, \p{0.16} the conformal cross-ratios \p{cr} become\footnote{The variables $z,\bar z$ are real and independent from each other in the case of Minkowski metric, and complex conjugate to each other in the Euclidean case \cite{Dolan:2004mu}. Similar variables have been used in the past for solving the superconformal Ward identities for four-point correlation functions \cite{partialNonRen,Dolan:2001tt,Heslop:2002hp,Dolan:2004mu}.}
\begin{align}\label{0.13}
u= \frac{x_{4}^2}{x_{1}^2} = (1-z)(1-\bar z)
\,,\qquad v=\frac{x_{41}^2}{x_{1}^2} = z\bar z\,,
\end{align}
and similarly for the $y-$variables:
\begin{align}\label{0.14}
 \frac{y_{12}^2y_{34}^2}{y_{13}^2 y_{24}^2} \ \to\ \frac{y_{4}^2}{y_{1}^2} = (1-w)(1-\bar w)
\,,\qquad  \frac{y_{23}^2y_{41}^2}{y_{13}^2 y_{24}^2}\ \to\ \frac{y_{41}^2}{y_{1}^2} = w\bar w \,.
\end{align}
Then, the product of the first four delta functions on the right-hand side of \p{0.1} reduces 
in the gauge \p{Kga} to
\begin{align}  \label{0.15}
(x_1^2 y_1^2 x_2^2 y_2^2)^2 \delta^{(4)}\lr{ \bar\xi' }\delta^{(4)} \lr{\epsilon} 
\delta^{(4)} \lr{ \bar\xi} \delta^{(4)} \lr{ \epsilon' y_{41} y_1^{-1}
 - x_{41} x_1^{-1} \epsilon'  }\,,
\end{align}
so that the integration over $\epsilon,\bar\xi,  \bar\xi'$ in \p{0.1} is trivial. To
integrate over $\epsilon'$ we introduce a parametrisation for its components
and use \p{0.13} and \p{0.14} to recast the argument of the last delta function in \p{0.15} in the form
\begin{align}
\epsilon' =  \left[\begin{array}{cc}\alpha  & \beta  \\  \gamma & \delta   \end{array}\right]  \qquad \Longrightarrow \qquad
 \epsilon' y_{41} y_1^{-1}
 - x_{41} x_1^{-1} \epsilon' = \left[\begin{array}{cc}\alpha (w-z) & \beta (\bar w-z)  \\  \gamma 
 (w-\bar z)& \delta (\bar w-\bar z)  \end{array}\right]\,.
\end{align}
Thus,  in the gauge  \p{0.13} and \p{0.14} the nilpotent invariant \p{0.1}  becomes
\begin{align}\label{0.20}
 \cI_n\vert_{\rho_1=\ldots=\rho_4=0}  \ \to \    \lr{x_1^2   x_2^2  y_1^2 y_2^2  }^2 (w-z)(\bar w-z) (w-\bar z)(\bar w-\bar z) \times \prod_{i=5}^n\delta^{(4)}(\rho_i)\,.
\end{align}
Let us now examine the expression for the rational factor $R$ from \p{eq:7}  in the same gauge. It is easy to check using \p{0.13} and \p{0.14} that it takes the following form \cite{Heslop:2002hp}
\begin{align}\label{0.21}
(x_{13}^2 x_{24}^2 x_{12}^2{x_{23}^2x_{34}^2x_{41}^2}) \times R(1,2,3,4) \ \to \  \lr{x_1^2   x_2^2  y_1^2 y_2^2  }^2 (w-z)(\bar w-z) (w-\bar z)(\bar w-\bar z) \,.    
\end{align}
Comparison of  \p{0.20} and \p{0.21} yields
\begin{align}\label{}
\cI_n\vert_{\rho_1=\ldots=\rho_4=0} =  (x_{13}^2 x_{24}^2 x_{12}^2{x_{23}^2x_{34}^2x_{41}^2}) \times R(1,2,3,4) \times \prod_{i=5}^n\delta^{(4)}(\rho_i)\,,
\end{align} 
exactly as announced in  \p{eq:6}.

\section{Numerical evaluation of the short-distance limit}\label{sec:three-loop-correlation function}

In this appendix, we study the asymptotic expansion of the Feynman integrals 
defined in \p{eq:14} and \p{eq:g+h} in the Euclidean region $x_1 \rightarrow \, x_2, \, x_3 \rightarrow \, x_4$. In terms of the conformal cross-ratios \p{cr}, this limit corresponds to $u\rightarrow 0$, $v\rightarrow 1$. As was explained in Section~4,
the advantage of this limit compared to the conventional short-distance  limit
$x_1\to x_2$, corresponding to $u\to 0$ with $v$ fixed, is that it allows us to 
single out the contribution to the OPE  \p{OPE-gen} of a couple of operators with 
the lowest spin and scaling dimension.

We start with the one-loop ladder function $g(1,2,3,4)$ and examine the
`three-star' integral in \p{MB-example}.  At the first step, we dimensionally regularise the integral 
and rewrite it using the standard  Feynman parameter technique:  
\begin{align}\label{1lmb}
I= \frac{1}{4 \pi^2} \, \int \frac{d^{4-2 \epsilon}x_5}{x^2_{15} x^2_{25} x^2_{35}}
 = & \frac14 {\Gamma(2 - \epsilon) \Gamma(1 + \epsilon)}  \int_0^1
\frac{dq_1 dq_2 dq_3 \, \delta(1 - q_1 - q_2 - q_3)}{(q_1 q_2
x_{12}^2 + q_1 q_3 x_{13}^2 + q_2 q_3 x_{23}^2)^{1+\epsilon}} \,.
\end{align}
Then, we follow the usual Mellin-Barnes  (MB) procedure for
rewriting the 
sum of terms in the denominator as a product at the cost of
introducing two new (contour) integrals with variables $z_1$ and
$z_2$ (with fixed real part and varying imaginary part)
\begin{align}
I = \frac{\Gamma(2-\epsilon)}{4  (x_{13}^2)^{1 + \epsilon}}
\int_{-i\infty}^{+i\infty} &\frac{dz_1 dz_2}{(2 \pi i)^2}  \Gamma(-z_1) \Gamma(-z_2) \Gamma(1 + \epsilon + z_1 + z_2) 
\left(\frac{x_{12}^2}{x_{13}^2}\right)^{z_2} 
\left(\frac{x_{23}^2}{x_{13}^2}\right)^{z_1}  \nonumber \\
 & \times\int_0^1 dq_1 dq_2 dq_3 \, \delta(1 - q_1 - q_2 - q_3) \,
q_1^{-(1+\epsilon+z_1)} q_2^{-(1+\epsilon+z_2)} q_3^{z_1+z_2}\,,
\end{align}
where the integration contours run along the imaginary axis to the left of the poles
of the first two $\Gamma-$function and to the right of the poles of the third
$\Gamma-$function.
Finally we perform the Feynman parameter integration (see for example~\cite{MB}
for more details) to get
\begin{align}
I=  \frac{\Gamma(2-\epsilon)}{4 \Gamma(1 - 2 \epsilon)
(x_{13}^2)^{1 + \epsilon}} \int_{-i\infty}^{+i\infty} & \frac{dz_1 dz_2}{(2 \pi i)^2 } \Gamma(-\epsilon-z_1) \Gamma(-z_1)
\Gamma(-\epsilon-z_2) \Gamma(-z_2) \times \nonumber  \\
\times &   \Gamma(1 + z_1 + z_2) \Gamma( 1 + \epsilon
+ z_1 + z_2) \left(\frac{x_{12}^2}{x_{13}^2}\right)^{z_2} 
\left(\frac{x_{23}^2}{x_{13}^2}\right)^{z_1}\,. 
\end{align}
At first sight, it may seem surprising that we use a dimensionally regularised form of the integral \p{1lmb}, given the fact that  it is finite for $\epsilon\to 0$
and hence conformally covariant. The reason for this can be understood in more complicated integrals, for example, the three-loop ladder (see, e.g., \cite{magic}). The MB parametrisation of the integral produces factors of $\Gamma(-2\ep)$ in the denominator, which are compensated by poles arising due to the integration over the $z_i$ variables. The result is of course finite and non-vanishing. The fact that this standard technique works in $D = 4 - 2 \epsilon$ does not
clash with conformal invariance precisely because the integrals are finite: We reduce a
conformal integral in four dimensions to a non-conformal three-point integral.
Although the latter is computed in $D = 4 - 2 \epsilon$, in the resulting finite expression  we can simply upgrade the $O(\epsilon^0)$ part of the
representation to a conformal four-point integral. The procedure leads to
a substantially reduced number of MB parameters as opposed
to what one would obtain starting from the four-point integrals.

Coming back to our example \p{1lmb}, the integral is finite as anticipated and we can safely put $\epsilon=0$, leading together with \p{MB-example} to 
\begin{equation}
g(1,2,3,4)  = - \frac{1}{4  x_{13}^2 x_{24}^2}
\int_{-i\infty}^{+i\infty} \frac{dz_1 dz_2}{(2 \pi i)^2} \, \left[\Gamma(-z_1) \Gamma(-z_2) \Gamma(1 + z_1 + z_2)\right]^2 \, 
u^{z_2} \, v^{z_1} \, .
\end{equation}
The MB.m package~\cite{Czakon}
may be used to evaluate this integral. Here it suggests
we  put the real parts of the integration contours to $\{-3/16,-3/8\}$ for\
$\{z_1,z_2\}$, respectively. For $u\to 0$ and $v\to 1$, we
 obtained with the help of the MBasymptotics.m package
\cite{Czakon2}
\begin{eqnarray}
 g(1,2,3,4)
& = & \frac{1}{x_{24}^4} \, \left( \frac{1}{4} \ln u \, - \, \frac{1}{2}
\, + \, O(u) \right) \, . \label{phi1as}
\end{eqnarray}
To match our Wick rotation conventions (positive Euclidean
signature) the output of the Mellin-Barnes software has to
be changed by a minus sign per loop order.

For the two-loop integral $h$, Eq.~\p{eq:g+h}, and the three-loop integrals $L,E,H$, Eq.~\p{eq:14}, there is no complete point-exchange symmetry, so each of the integrals entering \p{loop3F}
must be considered separately. To obtain an MB representation for $h(1,2;3,4)$ we send point 4 to
infinity upon which either subintegral becomes a three-star like \p{1lmb}, yielding two MB parameters each. For the three-loop ladder $L$ and the
easy three-loop integral $E$ we start with the outer subintegrals, the middle integral
is then a three-star as before so that we obtain representations with a total
of six parameters (see~(\ref{eq:14}) for the integral definitions).

The hard integral $H$ is much more nested. Let us send point 4
to infinity as before. We would prefer to start with the central integral which
becomes a three star, aiming at a 9-parameter representation. Unfortunately,
the MB.m package fails to find an integration contour for this
representation.\footnote{The same difficulty was encountered when studying
the pentabox in \cite{EKS2}. The hard and easy integrals both have a pentabox
subintegral; we were in fact surprised that the problem did not occur when
constructing the representation for the easy integral $E$ described above.}
Starting with the outer integrals and leaving the central integration to the end
produces a 12-parameter representation which we can run; its $O(\epsilon^0)$
part contains several integrals with up to nine parameters.

In the asymptotics
the number of parameters is smallest in the leading log part; as observed also
on pole parts of IR or UV divergent integrals the dimensionality of the MB
integrals increases towards the finite part. Yet we can obtain rather good
numerical precision because we compute only number coefficients and not
functions. After (for the hard integral rather heavy) application of the
barnesroutines.m package \cite{Czakon2}, the worst parameter count occurs for
the finite part of the hard integral, which in the  
$u \rightarrow 0, v \rightarrow 1$\ limit contains a few five-parameter
integrals. For up to four parameters the Cuhre routines in the Cuba library
\cite{cuba} allow excellent precision, but in this case we could not obtain
more than three significant digits.

  The coefficients that we find are either simple rationals like $3/2,\, 5/4$
or simple rationals times $\zeta(3)$ or $\zeta(5)$, and linear combinations
thereof. The denominators of the coefficients of the  subleading $\ln u$ divergences are powers of 2 wherever
the MBasymptotics.m package leads to analytic results. Using this as a guideline
it is very easy to guess exact results from the usually excellently precise
numerical output as long as there is no mixture of rational numbers and 
$\zeta-$values. 

In this way, we find the following results for asymptotic behaviour of
loop integrals entering \p{loop3F} as $u\to 0$ and $v\to 1$ (we display
the leading terms only)\footnote{We would like to acknowledge the generous help of Volodya Smirnov
in deriving the analytic expression for the constant term in the last relation.}
\begin{eqnarray}
x_{24}^4 \, h(1,2;3,4) & = & \frac{3}{8} \, \zeta(3) \, , \nonumber \\
x_{24}^4 \, h(1,3;2,4) = x_{24}^4 \, h(1,4;2,3) 
& = & \frac{1}{32} \, (\ln u)^2 -
\frac{3}{16} \, \ln u + \frac{3}{8} \, , \nonumber \\
x_{24}^4 \, L(1,2;3,4) & = & - \frac{5}{16} \, \zeta(5) \, ,\nonumber \\
x_{24}^4\,  L(1,3;2,4)= x_{24}^4 \, L(1,4;2,3)  & = & \frac{1}{384} \, (\ln u)^3 -
\frac{1}{32} \, (\ln u)^2 + \frac{5}{32} \, \ln u - \frac{5}{16} \, ,
\nonumber \\
x_{24}^4 \,  E(1,2;3,4) & = & \frac{1}{192} \, (\ln u)^3 -
\frac{3}{64} \, (\ln u)^2 + \frac{5}{32} \, \ln u  \nonumber \\
&& -  \frac{5}{16} \,
\zeta(5) + \frac{3}{32} \, \zeta(3) - \frac{5}{32}   \, , \nonumber \\
x_{24}^4 \,  E(1,3;2,4)=x_{24}^4 \, E(1,4;2,3) & = &
\frac{3}{32} \, \zeta(3) \, \ln u - \frac{3}{16} \, \zeta(3) \, , \nonumber \\
x_{24}^4 \,  H(1,2;3,4) & = & \frac{3}{16} \, \zeta(3) \, \ln u
- \frac{3}{16} \, \zeta(3) \, , \nonumber \\
x_{24}^4 \,  H(1,3;2,4)=x_{24}^4 \, H(1,4;2,3)  & = & \frac{1}{192} \, (\ln u)^3
- \frac{1}{16} \, (\ln u)^2 + \frac{9}{32} \, \ln u +\frac14 \zeta(3)- \frac{7}{16} \notag
\\{}
\, . 
\label{Integrals}
\end{eqnarray}
Using these formulae we can find the short-distance limit of the loop
corrections (\ref{intriLoops}) to the four-point correlation function displayed in \p{allGs}.


\begin{thebibliography}{99}

\bibitem{123}
  J.~Maldacena, {\it Adv.~Theor.~Math.~Phys.} {\bf 2} (1998) 231
  [arXiv:hep-th/9711200];
  S.~Gubser, I.~Klebanov and A.~Polyakov, {\it Phys.~Lett.} {\bf B428}
  (1998) 105 [arXiv:hep-th/9802109];
  E.~Witten, {\it Adv.~Theor.~Math.~Phys.} {\bf 2} (1998) 253
  [arXiv:hep-th/9802150].

\bibitem{oneTwo}
  B.~Eden, P.~S.~Howe, C.~Schubert, E.~Sokatchev and P.~West,
  Nucl.\ Phys.\  {\bf B557 } (1999)  355-379
  [hep-th/9811172];
  Phys.\ Lett.\  {\bf B466 } (1999)  20-26
  [hep-th/9906051];
  F.~Gonzalez-Rey, I.~Y.~Park and K.~Schalm,
  Phys.\ Lett.\  {\bf B448 } (1999)  37-40
  [hep-th/9811155].
  
  
  \bibitem{ESS}
   B.~Eden, C.~Schubert and E.~Sokatchev,
  Phys.\ Lett.\  {\bf B482 } (2000)  309-314.
  [hep-th/0003096];
  B.~Eden, C.~Schubert and E.~Sokatchev, unpublished.
  
  \bibitem{Rome}
   M.~Bianchi, S.~Kovacs, G.~Rossi and Y.~Stanev,
  Nucl.\ Phys.\  {\bf B584 } (2000)  216-232
  [hep-th/0003203].

\bibitem{partialNonRen}
  B.~Eden, A.~C.~Petkou, C.~Schubert and E.~Sokatchev,
  Nucl.\ Phys.\  {\bf B607 } (2001)  191-212
  [hep-th/0009106].

\bibitem{EHW}
  B.~Eden, P.~S.~Howe and P.~C.~West,
  Phys.\ Lett.\  {\bf B463 } (1999) 19-26
  [hep-th/9905085].

\bibitem{Heslop:2001dr}
  P.~J.~Heslop, P.~S.~Howe,
  Phys.\ Lett.\  {\bf B516 } (2001)  367-375.
  [hep-th/0106238];
  P.~J.~Heslop, P.~S.~Howe,
  Nucl.\ Phys.\  {\bf B626 } (2002)  265-286.
  [hep-th/0107212].

\bibitem{Heslop:2003xu}
  P.~J.~Heslop and P.~S.~Howe,
  JHEP {\bf 0401} (2004) 058
  [arXiv:hep-th/0307210].
 
\bibitem{Heslop:2002hp}
  P.~J.~Heslop, P.~S.~Howe,
  JHEP {\bf 0301 } (2003)  043.
  [hep-th/0211252].

\bibitem{Dolan:2004mu}
  F.~A.~Dolan, L.~Gallot and E.~Sokatchev,
  JHEP {\bf 0409} (2004) 056
  [arXiv:hep-th/0405180].
 

\bibitem{Dolan:2001tt}
  F.~A.~Dolan and H.~Osborn,
  Nucl.\ Phys.\  B {\bf 629} (2002) 3
  [arXiv:hep-th/0112251].


 

\bibitem{aldMald}
  L.~Alday and J.~Maldacena, {\it JHEP} {\bf 0706} (2007) 064
  [arXiv:0705.0303 [hep-th]].

\bibitem{dual11}
  J.~M.~Drummond, G.~P.~Korchemsky and E.~Sokatchev,
  {\it Nucl.~Phys.} {\bf B795} (2008) 385 [arXiv:0707.0243 [hep-th]].

\bibitem{dual12}
  A.~Brandhuber, P.~Heslop and G.~Travaglini,
  {\it Nucl.~Phys.} {\bf B794} (2008) 231 [arXiv:0707.1153 [hep-th]].

\bibitem{dual21}
  J.~Drummond, J.~Henn, G.~Korchemsky and E.~Sokatchev, {\it Nucl.~Phys.}
  {\bf B815} (2009) 142 [arXiv:0803.1466 [hep-th]].

\bibitem{dual22}
  Z.~Bern, L.~Dixon, D.~Kosower, R.~Roiban, M.~Spradlin, C.~Vergu and
  A.~Volovich, {\it Phys.~Rev.} {\bf D78} (2008) 045007
  [arXiv:0803.1465 [hep-th]].

\bibitem{Anastasiou:2009kna}
  C.~Anastasiou, A.~Brandhuber, P.~Heslop, V.~V.~Khoze, B.~Spence, G.~Travaglini,
  JHEP {\bf 0905 } (2009)  115.
  [arXiv:0902.2245 [hep-th]].


\bibitem{DelDuca:2010zg}
  V.~Del Duca, C.~Duhr, V.~A.~Smirnov,
  JHEP {\bf 1005 } (2010)  084.
  [arXiv:1003.1702 [hep-th]].

\bibitem{DelDuca:2010zp}
  V.~Del Duca, C.~Duhr, V.~A.~Smirnov,
  JHEP {\bf 1009 } (2010)  015.
  [arXiv:1006.4127 [hep-th]].

\bibitem{spradGonch}
  A.~Goncharov, M.~Spradlin, C.~Vergu and A.~Volovich,
  {\it Phys.~Rev.~Lett.}  {\bf 105} (2010) 151605
  [arXiv:1006.5703 [hep-th]].

\bibitem{Heslop:2010kq}
  P.~Heslop, V.~V.~Khoze,
  JHEP {\bf 1011 } (2010)  035.
  [arXiv:1007.1805 [hep-th]].


\bibitem{annecySuperspace}
  J.~M.~Drummond, J.~Henn, G.~P.~Korchemsky and E.~Sokatchev,
  Nucl.\ Phys.\  {\bf B828 } (2010)  317-374
  [arXiv:0807.1095 [hep-th]].

\bibitem{BM}
  N.~Berkovits and J.~Maldacena,
  JHEP {\bf 0809} (2008) 062
  [arXiv:0807.3196];
\\
  N.~Beisert, R.~Ricci, A.~A.~Tseytlin and M.~Wolf,
  Phys.\ Rev.\  D {\bf 78}, 126004 (2008)
  [arXiv:0807.3228].

\bibitem{Brandhuber:2008pf}
  A.~Brandhuber, P.~Heslop and G.~Travaglini,
  Phys.\ Rev.\  {\bf D78 } (2008)  125005.
  [arXiv:0807.4097 [hep-th]].

\bibitem{nima1}
  N.~Arkani-Hamed, J.~L.~Bourjaily, F.~Cachazo, S.~Caron-Huot and J.~Trnka,
  JHEP {\bf 1101} (2011) 041 [arXiv:1008.2958 [hep-th]].

\bibitem{nima2}
  N.~Arkani-Hamed, J.~L.~Bourjaily, F.~Cachazo and J.~Trnka,
  arXiv:1012.6032 [hep-th].

\bibitem{marksAndSparks}
  L.~J.~Mason and D.~Skinner,
  JHEP {\bf 1012} (2010) 018 [arXiv:1009.2225 [hep-th]].



\bibitem{simon}
  S.~Caron-Huot,
  arXiv:1010.1167 [hep-th].

\bibitem{belitsky}
  A.~V.~Belitsky, G.~P.~Korchemsky and E.~Sokatchev,
 arXiv:1103.3008 [hep-th].


\bibitem{AEMKS}
  L.~F.~Alday, B.~Eden, G.~P.~Korchemsky, J.~Maldacena and E.~Sokatchev,
  arXiv:1007.3243 [hep-th].


\bibitem{EKS2}
  B.~Eden, G.~P.~Korchemsky and E.~Sokatchev,
  arXiv:1007.3246 [hep-th].

\bibitem{EKS3}
  B.~Eden, G.~P.~Korchemsky and E.~Sokatchev,
  arXiv:1009.2488 [hep-th].

\bibitem{paper1}
  B.~Eden, P.~Heslop, G.~P.~Korchemsky and E.~Sokatchev,
  arXiv:1103.3714 [hep-th].

\bibitem{paper2}
  B.~Eden, P.~Heslop, G.~P.~Korchemsky and E.~Sokatchev,
  arXiv:1103.4353 [hep-th].


\bibitem{mscorrelationfunctions}
  T.~Adamo, M.~Bullimore, L.~Mason, D.~Skinner,
  [arXiv:1103.4119 [hep-th]].


\bibitem{ADS} E. D'Hoker, D. Z. Freedman, S. D. Mathur, A. Matusis and L. Rastelli,
Nucl.Phys. {\bf B562} (1999) 353-394, hep-th/9903196;
 E. D'Hoker, S. D. Mathur, A. Matusis and L. Rastelli,
Nucl.Phys. {\bf B589} (2000) 38-74, hep-th/9911222;
  G. Arutyunov and S. Frolov,
Phys.Rev. {\bf D62} (2000) 064016, hep-th/0002170;
G.~Arutyunov, F.~A.~Dolan, H.~Osborn, E.~Sokatchev,
  Nucl.\ Phys.\  {\bf B665 } (2003)  273-324.
  [hep-th/0212116].
  
  
\bibitem{paulN4}
  G.~G.~Hartwell and P.~S.~Howe,
  Int.\ J.\ Mod.\ Phys.\  {\bf A10 } (1995) 3901-3920
  [hep-th/9412147];
  Class.\ Quant.\ Grav.\  {\bf 12 } (1995) 1823-1880.

\bibitem{bern42}
  Z.~Bern, L.~J.~Dixon, V.~A.~Smirnov,
  Phys.\ Rev.\  {\bf D72 } (2005)  085001
  [hep-th/0505205].



\bibitem{4loopMHV}
  Z.~Bern, M.~Czakon, L.~J.~Dixon, D.~A.~Kosower, V.~A.~Smirnov,
  Phys.\ Rev.\  {\bf D75 } (2007)  085010.
  [hep-th/0610248].

\bibitem{5loops}
  Z.~Bern, J.~J.~M.~Carrasco, H.~Johansson, D.~A.~Kosower,
  Phys.\ Rev.\  {\bf D76 } (2007)  125020.
  [arXiv:0705.1864 [hep-th]].

\bibitem{klovUs}
  A.~V.~Kotikov, L.~N.~Lipatov, A.~I.~Onishchenko, V.~N.~Velizhanin,
  Phys.\ Lett.\  {\bf B595 } (2004)  521-529 
  [hep-th/0404092];
  B.~Eden, C.~Jarczak, E.~Sokatchev,
  Nucl.\ Phys.\  {\bf B712 } (2005)  157-195 
  [hep-th/0409009].

\bibitem{Andrianopoli:1999vr}
  L.~Andrianopoli, S.~Ferrara, E.~Sokatchev, B.~Zupnik,
  Adv.\ Theor.\ Math.\ Phys.\  {\bf 4 } (2000)  1149-1197.
  [hep-th/9912007].
 
  
\bibitem{GIKOS}
  A.~Galperin, E.~Ivanov, S.~Kalitsyn, V.~Ogievetsky and E.~Sokatchev,
  Class.\ Quant.\ Grav.\  {\bf 1 } (1984)  469-498;
  A.~Galperin, E.~Ivanov, V.~Ogievetsky and E.~Sokatchev,
  Class.\ Quant.\ Grav.\  {\bf 2 } (1985)  601;
  Class.\ Quant.\ Grav.\  {\bf 2 } (1985)  617;\\
  A.~S.~Galperin, E.~A.~Ivanov, V.~I.~Ogievetsky, E.~S.~Sokatchev,
  ``Harmonic superspace,''
  Cambridge, UK: Univ. Pr. (2001) 306 p.
  
\bibitem{Eden:2012tu}
  B.~Eden, P.~Heslop, G.~P.~Korchemsky and E.~Sokatchev,
  arXiv:1201.5329 [hep-th].


\bibitem{Anastasiou:2003kj}
  C.~Anastasiou, Z.~Bern, L.~J.~Dixon, D.~A.~Kosower,
  Phys.\ Rev.\ Lett.\  {\bf 91 } (2003)  251602.
  [hep-th/0309040].

  
\bibitem{magic}
  J.~M.~Drummond, J.~Henn, V.~A.~Smirnov and E.~Sokatchev,
  JHEP {\bf 0701 } (2007)  064
  [hep-th/0607160].

\bibitem{Kon12}
  D.~Anselmi, M.~T.~Grisaru, A.~Johansen,
  Nucl.\ Phys.\  {\bf B491 } (1997)  221-248.
  [hep-th/9601023].

\bibitem{davyPhi1}
  N.~I.~Usyukina and A.~I.~Davydychev,
  Phys.\ Lett.\  B {\bf 298} (1993) 363;  
  N.~I.~Usyukina and A.~I.~Davydychev,
  Phys.\ Lett.\  B {\bf 305} (1993) 136.

\bibitem{MB}
  V.~A.~Smirnov,
  ``Feynman integral calculus,''
  {\it  Berlin, Germany: Springer (2006)} 283~pp.

\bibitem{multigraph}
See \href{http://www.mathe2.uni-bayreuth.de/markus/multigraphs.html}{http://www.mathe2.uni-bayreuth.de/markus/multigraphs.html}

\bibitem{Cachazo:2008dx}
  F.~Cachazo, D.~Skinner,
  [arXiv:0801.4574 [hep-th]].
  
\bibitem{Belitsky:2003sh}
  A.~V.~Belitsky, S.~E.~Derkachov, G.~P.~Korchemsky, A.~N.~Manashov,
  Phys.\ Rev.\  {\bf D70 } (2004)  045021.
  [hep-th/0311104].

\bibitem{Arutyunov:2003ad}
  G.~Arutyunov, S.~Penati, A.~Santambrogio, E.~Sokatchev,
  Nucl.\ Phys.\  {\bf B670 } (2003)  103-147.
  [hep-th/0305060].
  
  
\bibitem{Czakon}
  M.~Czakon,
  Comput.\ Phys.\ Commun.\  {\bf 175 } (2006)  559-571
  [hep-ph/0511200].

\bibitem{Czakon2}
see \href{http://projects.hepforge.org/mbtools/}{http://projects.hepforge.org/mbtools/}

\bibitem{cuba}
  T.~Hahn,
  Comput.\ Phys.\ Commun.\  {\bf 168 } (2005)  78-95.
  [hep-ph/0404043].

\end{thebibliography}
 \end{document}